\documentclass[11pt]{article}
\usepackage[margin=1in]{geometry}
\usepackage{hyperref}
\usepackage{url}
\usepackage{float}
\usepackage{xcolor}
\usepackage{booktabs}
\usepackage{tabularx}
\usepackage{graphicx}
\usepackage{caption}
\usepackage{subcaption}
\usepackage[protrusion=false,expansion=false]{microtype}
\usepackage{parskip}
\usepackage{titlesec}
\usepackage{enumitem}
\usepackage{amsmath}
\usepackage{tikz}
\usetikzlibrary{arrows.meta, positioning, backgrounds, fit, decorations.markings}
 
\definecolor{codebg}{HTML}{F6F8FA}
\definecolor{diffadd}{HTML}{E6FFED}
\definecolor{diffdel}{HTML}{FFEEF0}
\definecolor{codegreen}{HTML}{22863A}
\definecolor{codepurple}{HTML}{6F42C1}
\definecolor{codeblue}{HTML}{005CC5}
\definecolor{codegray}{HTML}{6A737D}
\definecolor{colBorder}  {HTML}{1A252F}
\definecolor{colExternal}{HTML}{D6EAF8}
\definecolor{colChannel} {HTML}{AED6F1}
\definecolor{colGateway} {HTML}{A9DFBF}
\definecolor{colAgent}   {HTML}{FAD7A0}
\definecolor{colNode}    {HTML}{F1948A}
\definecolor{colPlugin}  {HTML}{F9E79F}
\definecolor{colFlow}    {HTML}{1A5276}
\definecolor{colAttack}  {HTML}{C0392B}
\definecolor{colKC}      {HTML}{7B241C}
\definecolor{colTrust}   {HTML}{808B96}
\definecolor{colSubBg}   {HTML}{FDFEFE}
\definecolor{cIA}  {HTML}{2E86C1}
\definecolor{cCM}  {HTML}{D35400}
\definecolor{cEX}  {HTML}{1A7A4A}
\definecolor{cCA}  {HTML}{8E44AD}
\definecolor{cPE}  {HTML}{B7950B}
\definecolor{cIM}  {HTML}{C0392B}
\definecolor{cBG}  {HTML}{F4F6F7}
\definecolor{cSUB} {HTML}{EBF5FB}
\definecolor{cBORD}{HTML}{2C3E50}
\tikzset{
  tacticH/.style={draw=#1, fill=#1, text=white, rounded corners=3pt,
    minimum width=3.65cm, minimum height=0.72cm,
    font=\small\bfseries\sffamily, align=center},
  techcard/.style={draw=#1!60, fill=#1!12, line width=0.6pt,
    minimum width=3.65cm, minimum height=0.52cm, rounded corners=2pt,
    font=\scriptsize\bfseries\sffamily\color{cBORD},
    align=flush left, inner xsep=5pt, inner ysep=2pt},
  subcard/.style={draw=#1!40, fill=cSUB, line width=0.5pt,
    minimum width=3.35cm, minimum height=0.46cm, rounded corners=1.5pt,
    font=\tiny\sffamily\color{cBORD!80},
    align=flush left, inner xsep=4pt, inner ysep=2pt},
}
\definecolor{colKCa}{HTML}{2E86C1}
\definecolor{colKCb}{HTML}{D35400}
\definecolor{colKCd}{HTML}{8E44AD}
\definecolor{colKCe}{HTML}{C0392B}
\definecolor{colSwimkc}{HTML}{F2F3F4}
\tikzset{
  stepbox/.style={draw=black!30, fill=white, line width=0.7pt,
    rounded corners=4pt, align=center, font=\scriptsize, inner sep=4pt},
  vulntag/.style={draw=red!70!black, fill=red!8, line width=0.6pt,
    rounded corners=2pt, inner sep=2pt,
    font=\tiny\bfseries\color{red!70!black}, align=center},
  layertag/.style={draw=black!25, fill=green!8, line width=0.5pt,
    rounded corners=2pt, inner sep=2pt,
    font=\tiny\color{black!60}, align=center},
}
\definecolor{colKC1}{HTML}{2E86C1}
\definecolor{colKC2}{HTML}{D35400}
\definecolor{colKC3}{HTML}{8E44AD}
\definecolor{colKC4}{HTML}{C0392B}
\definecolor{colSwim}{HTML}{F2F3F4}
\definecolor{colVuln}{HTML}{FDEDEC}
\definecolor{colLayer}{HTML}{D5F5E3}

\usepackage{listings}
\usepackage{xcolor}  % already loaded

\definecolor{codebg}{rgb}{0.97, 0.97, 0.97}

\definecolor{kwcolor}{rgb}{0.0, 0.0, 0.8}      % blue: export, function, const, if, return
\definecolor{typecolor}{rgb}{0.13, 0.55, 0.13}  % green: string, number, null, undefined, true, false
\definecolor{strcolor}{rgb}{0.7, 0.1, 0.1}      % red/brown: "wildcard", "id", "name"
\definecolor{commentcolor}{rgb}{0.4, 0.4, 0.4}  % gray: comments

\lstdefinelanguage{TypeScript}{
  keywords={export, function, const, let, var, if, return, true, false, 
             typeof, instanceof, void, new, class, extends, import, from,
             as, type, interface, enum, for, of, in, while, break, continue},
  keywordstyle=\color{kwcolor},
  morekeywords=[2]{string, number, null, undefined, boolean, Array, void},
  keywordstyle=[2]\color{typecolor},
  sensitive=true,
  morestring=[b]",
  morestring=[b]',
  morestring=[b]`,
  stringstyle=\color{strcolor},
  morecomment=[l]{//},
  commentstyle=\color{commentcolor}\itshape,
}

\usepackage{newfloat}
\DeclareFloatingEnvironment[
  fileext=lol,
  listname={List of Listings},
  name=Listing,
  placement=tbp,
]{listing}

\titleformat{\section}{\large\bfseries}{\thesection}{1em}{}
\titleformat{\subsection}{\normalsize\bfseries}{\thesubsection}{1em}{}
\titleformat{\subsubsection}{\normalsize\itshape\bfseries}{\thesubsubsection}{1em}{}

\begin{document}

\title{\textbf{A Security Analysis of the OpenClaw AI Agent Framework}}

\author{
  Surada Suwansathit\\
    \textit{SUCCESS Lab}\\
  \textit{Texas A\&M University}\\
  \texttt{surada@tamu.edu}
  \and
  Yuxuan Zhang\\
      \textit{SUCCESS Lab}\\
  \textit{Texas A\&M University}\\
  \texttt{yuz516@tamu.edu}
  \and
  Guofei Gu\\
      \textit{SUCCESS Lab}\\
  \textit{Texas A\&M University}\\
  \texttt{guofei@cse.tamu.edu}
}

\date{April 2026}
\maketitle
% === ABSTRACT ===
\begin{abstract}
AI agent frameworks that connect large language model (LLM) reasoning to host
execution surfaces---shell, filesystem, containers, browser automation, and
messaging platforms---introduce a class of security challenges that differs
structurally from those of conventional software. We present a systematic
\emph{security taxonomy} of 470 advisories filed against OpenClaw
\cite{openclaw-repo}, an open-source AI agent runtime, organizing the corpus
by \emph{architectural layer} and \emph{trust-violation type}. Our taxonomy reveals that vulnerabilities cluster along two
orthogonal axes: (1)~the \emph{system axis}, which reflects where in the
architecture a vulnerable operation occurs (exec policy, gateway, channel
adapters, sandbox, browser, plugin/skill, agent/prompt layers); and
(2)~the \emph{attack axis}, which reflects the adversarial technique applied
(identity spoofing, policy bypass, cross-layer composition, prompt injection,
supply-chain trust escalation).
Grounding our taxonomy in patch-differential evidence, we derive three
principal findings. First, three independently Moderate- or High-severity advisories \cite{ghsa-g8p2, ghsa-gv46} exist within the Gateway and Node-Host subsystems. By mapping these through the Delivery, Exploitation, and Command-and-Control stages of the Cyber Kill Chain, they compose into a complete unauthenticated remote code execution path from an LLM tool call to the host process. Second, the exec allowlist
\cite{ghsa-9868, ghsa-gwqp, ghsa-3c6h}, which is the framework's primary
command-filtering mechanism, embeds a closed-world assumption that command
identity is recoverable by lexical parsing. This assumption is invalidated by line
continuation, busybox multiplexing, and GNU long-option abbreviation in
independent and non-overlapping ways. Third, a malicious skill distributed
through the plugin channel \cite{ghsa-yahoofinance} executed a two-stage dropper
entirely within the LLM context, bypassing the exec pipeline entirely and
illustrating that the skill distribution surface constitutes an attack vector
outside any runtime policy primitive. Across all categories, the dominant
structural pattern is per-layer, per-call-site trust enforcement rather than
unified policy boundaries---a design property that makes cross-layer composition
attacks systematically resistant to layer-local remediation. 
% We conclude by proposing defense strategies mapped to each taxonomy branch.
\end{abstract}

% === INTRODUCTION ===
\section{Introduction}
\label{sec:introduction}

\paragraph{AI agent frameworks as an execution surface.}
The deployment of large language models as autonomous agents---systems that
perceive external input, reason over it, and produce actions with real-world
effects---introduces a qualitatively different security problem from that of
conventional application software. In a traditional application, the code
determines behavior; an attacker who cannot modify the code is largely confined
to exploiting memory safety errors or logic bugs in well-defined input handlers.
In an AI agent framework, the model's output is itself a control signal: a tool
call emitted by the model instructs the runtime to execute a shell command, read
or write a file, navigate a browser, or deliver a message across a messaging
platform. The attack surface therefore includes not only the runtime's
implementation correctness but also the model's susceptibility to adversarial
influence through any data path that reaches its context window.

OpenClaw \cite{openclaw-repo} is a representative instance of this architecture.
The framework exposes a distributed agent runtime connecting LLM inference to
more than fifteen external surfaces through a layered Gateway--Node-Host design.
Its four principal subsystems---a Gateway control plane, a Node-Host privileged
execution process, an embedded agent runner, and a set of channel adapters
for messaging platforms---interact over WebSocket connections with
authentication and trust decisions scattered across each layer. The framework's
rapid adoption---exceeding 200,000 GitHub stars within weeks of its January
2026 relaunch under the OpenClaw name---made it an unusually high-visibility
target for security researchers during the precise window when it lacked a
mature disclosure process.

\paragraph{The need for a systematic taxonomy.}
Prior work on AI agent security \cite{perez2022prompt, greshake2023youve,
rando2024gradient} characterizes individual attack techniques---prompt injection,
indirect injection, model extraction---without a unifying model that maps
attacks to the specific architectural layer they exploit. A corpus as large and
structurally varied as OpenClaw's 470 advisories cannot be understood through a
simple list; its security implications emerge from the \emph{relationships}
between vulnerabilities across layers. We therefore organize our analysis along
two independent axes. The \textbf{system axis} (§\ref{sec:taxonomy-system})
classifies each vulnerability by the architectural component in which the
vulnerable operation occurs. The \textbf{attack axis}
(§\ref{sec:taxonomy}) classifies each vulnerability by the adversarial
technique, mapped to the Cyber Kill Chain \cite{lockheed-killchain} where
applicable. Together the two axes form a two-dimensional taxonomy that exposes
which architectural layers are susceptible to which technique classes, and
where defenses should be positioned.

\paragraph{Contributions.}
This paper makes the following contributions:

\begin{enumerate}[noitemsep]
  \item \textbf{A two-axis security taxonomy} of AI agent framework
    vulnerabilities, instantiated on the full 470-advisory OpenClaw corpus,
    organized by \emph{architectural layer} (system axis) and
    \emph{adversarial technique} (attack axis), with advisory citations
    supporting each classification (§\ref{sec:taxonomy}).

  \item \textbf{An OpenClaw-specific kill chain} that adapts five MITRE
    ATT\&CK tactics to the personal AI agent context and introduces
    Context Manipulation as a novel stage with no analog in traditional
    intrusion frameworks, reflecting the unique role of the LLM reasoning
    layer as an attack surface (§\ref{sec:taxonomy-killchain}).

  % \item \textbf{A structural analysis} of the exec allowlist
  %   \cite{ghsa-9868, ghsa-gwqp, ghsa-3c6h} demonstrating that its security
  %   property rests on a closed-world assumption about command-text parseability
  %   that is invalidated by at least three independent OS- and shell-level
  %   mechanisms (§\ref{sec:vulnanalysis}).

  % \item \textbf{A supply-chain trust case study} of the malicious
  %   \texttt{yahoofinance} skill \cite{ghsa-yahoofinance}, which bypassed the
  %   exec pipeline entirely and demonstrates that the skill distribution surface
  %   constitutes an attack vector not addressed by any runtime policy mechanism
  %   (§\ref{sec:plugin-skill}).
  \item \textbf{A multi-layer vulnerability analysis} of the 470-advisory OpenClaw corpus, providing a systematic mapping of empirical attack data to the ten architectural layers defined in our taxonomy. This analysis identifies recurring design flaws—ranging from identity mutability at the Channel Input Interface to lexical parsing failures in the Exec Policy Engine—demonstrating how decentralized trust boundaries enable complex, cross-layer exploitation chains (\ref{sec:multi-layer vulnerability analysis}).

  % \item \textbf{A defense framework} mapping recommended mitigations to each
  %   taxonomy branch, derived from the architectural root causes identified
  %   across the corpus (§\ref{sec:defenses}).

  % \item \textbf{A structural finding:} the codebase consistently enforces
  %   trust at individual call sites within each layer rather than at unified
  %   inter-layer boundaries, a design property that makes cross-layer
  %   composition attacks resistant to layer-local remediation.
\end{enumerate}

% \paragraph{Paper organization.}
% Section~\ref{sec:background} describes OpenClaw's architecture and the four
% principal subsystems that define its trust model, including both the
% \emph{system perspective} (how components are connected) and the
% \emph{attack perspective} (where trust boundaries can be violated).
% Section~\ref{sec:taxonomy} presents our two-axis taxonomy and maps all 190
% advisories to its branches.
% Section~\ref{sec:statistics} presents aggregate statistics over the full
% advisory corpus and the two disclosure waves.
% Section~\ref{sec:plugin-skill} covers the plugin and skill system with
% extended treatment of the malicious \texttt{yahoofinance} skill.
% Section~\ref{sec:defenses} points out possible future direction for defense frameworks.
% We conclude in Section~\ref{sec:conclusion}.
\paragraph{Paper organization.}
Section~\ref{sec:background} describes OpenClaw’s architecture and the principal subsystems defining its trust model. Section~\ref{sec:statistics} provides a high-level overview of the 470-advisory corpus and disclosure statistics. Section~\ref{sec:taxonomy} introduces our two-axis taxonomy, defining ten architectural layers and seven adversarial techniques. Section~\ref{sec:multi-layer vulnerability analysis} presents a structural vulnerability analysis, mapping empirical audit data to the taxonomy layers to identify architectural root causes. Section~\ref{sec:defenses} outlines potential defense strategies and mitigation directions aligned with each architectural layer of the proposed taxonomy. We conclude in Section~\ref{sec:conclusion}.
% ══════════════════════════════════════════════════════════════════════════════
\section{Modeling System Architecture}
\label{sec:background}
% ══════════════════════════════════════════════════════════════════════════════

\begin{figure}
    \centering
    \includegraphics[width=1\linewidth]{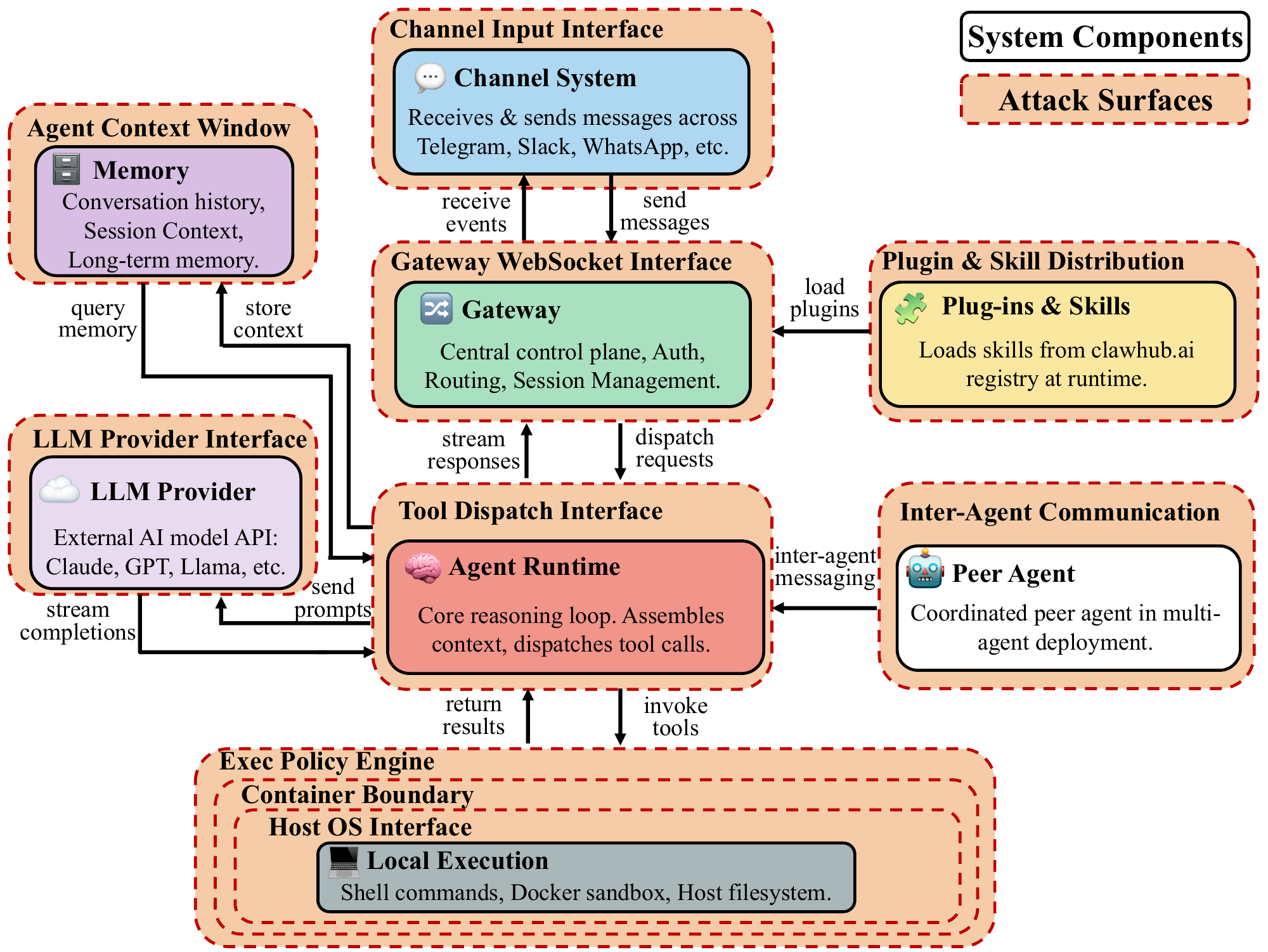}
    \caption{OpenClaw system architecture with attack surfaces mapped to each component. Solid boxes represent system components; dashed orange regions denote attack surfaces from the taxonomy in Table~\ref{fig:taxonomy-matrix}.}
    \label{fig:arch}
\end{figure}

OpenClaw \cite{openclaw-repo} is an open-source autonomous AI agent framework
that connects large language model inference to real-world execution surfaces:
shell command execution, file system access, browser automation, Docker container
management, and a wide array of third-party messaging platforms. Released in
November 2025 under the name Clawdbot, renamed to Moltbot in January 2026
following a trademark dispute, and rebranded to OpenClaw on January 29, 2026,
the project accumulated over 100,000 GitHub stars within weeks of its initial
viral distribution. Its architecture is organized around seven interacting
components (Figure~\ref{fig:arch}): a Channel System, a central Gateway, a
Plug-ins \& Skills System, an Agent Runtime, a Memory \& Knowledge System, an
LLM Provider, and a Local Execution environment.

\subsection{System Components}

\paragraph{Channel System.}
The Channel System (\textit{src/telegram/}, \textit{src/discord/},
\textit{src/slack/}, etc)) bridges external messaging platforms to the rest of the framework.
Each adapter polls or receives webhook events, authorizes the sender against
per-channel allowlists in \textit{src/channels/allow-from.ts}, computes a
canonical session key, and dispatches an inbound message to the Gateway's
command queue. Outbound responses follow the same path in reverse.

\paragraph{Gateway.}
The Gateway (\textit{src/gateway/}) is the central control plane and message
broker. It binds an HTTP/WebSocket server and is responsible for
authenticating and multiplexing all inbound connections from channel adapters,
the Agent Runtime, Local Execution processes, and CLI operators. It maintains a
\texttt{NodeRegistry} of connected Local Execution sessions, an
\texttt{ExecApprovalManager} that serializes pending command-approval requests,
and a per-lane \texttt{CommandQueue} that serializes concurrent messages
destined for the same session. The Gateway routes \texttt{node.invoke} frames
from the Agent Runtime to the appropriate Local Execution process and brokers
all AI agent runs.

\paragraph{Plug-ins \& Skills System.}
The Plug-ins \& Skills System manages the loading and execution of third-party
skills from the \texttt{clawhub.ai} registry and local plugin directories
(\textit{src/plugins/}). Skills are loaded into the agent's context window at
session start via \texttt{SKILL.md} instruction files, extending the agent's
capabilities at runtime. This component operates at operator-level trust: a
skill loaded by the system is treated as a trusted instruction source by the
Agent Runtime.

\paragraph{Agent Runtime.}
The Agent Runtime (\textit{src/agents/}) encapsulates the LLM reasoning loop,
tool dispatch, and Docker sandbox management. The entry point
\texttt{runEmbeddedPiAgent} (\textit{src/agents/pi-embedded-runner/run.ts})
resolves auth profiles, selects a model, and submits turns to the LLM Provider
in an attempt loop with failover. Tool calls emitted by the model are
intercepted by handlers in
\textit{src/agents/pi-embedded-subscribe/handlers/tools.ts} and dispatched
either in-process (file reads, web fetches) or forwarded to the Gateway as
\texttt{node.invoke} frames for host-side execution via Local Execution.

\paragraph{Memory \& Knowledge System.}
The Memory \& Knowledge System manages session history, long-term context, and
bootstrap files loaded at the start of each agent turn. The embedded runner
prepends \texttt{CLAUDE.md}, loaded skill instructions, and prior conversation
history into the LLM context window before each model call, giving the agent
persistent memory across turns within a session.

\paragraph{LLM Provider.}
The LLM Provider is the external AI model API (Claude, GPT, Llama, or any
locally-hosted model) that receives assembled prompts from the Agent Runtime
and streams completions back. The provider interface is abstracted in
\textit{src/agents/}, allowing model substitution without changes to the agent
runtime logic.

\paragraph{Local Execution.}
The Local Execution environment (\textit{src/node-host/}) runs on the
end-user's machine as a privileged process. It connects to the Gateway over
WebSocket with \textit{role="node"} and waits for \texttt{node.invoke} frames.
The core dispatch loop in \textit{src/node-host/invoke.ts} routes each command
through a three-phase exec policy pipeline: lexical allowlist evaluation,
approval state lookup, and execution. Sandboxed tool calls execute inside a
Docker container via \texttt{docker exec}; unsandboxed calls run directly on
the host shell with full filesystem and process access.

% ══════════════════════════════════════════════════════════════════════════════
\section{Corpus Overview}
\label{sec:statistics}
% ══════════════════════════════════════════════════════════════════════════════

Between January~31 and April~15, 2026, 470 security advisories were filed
against OpenClaw across three disclosure waves. The first wave (Jan~31--Feb~16)
comprised a coordinated 73-advisory audit. A second, accelerated wave
(Feb~17--28) added 150 advisories while patches from the first wave were still
being merged, more than doubling the corpus in under two weeks. A third wave
(Mar--Apr~15) contributed a further 247 advisories as the project's security
disclosure process matured and the researcher community broadened.
Table~\ref{tab:corpus-comparison} shows how the distribution across attack
surfaces evolved between the initial audit cutoff and the full corpus.

\begin{figure}
    \centering
    \includegraphics[width=0.9\linewidth]{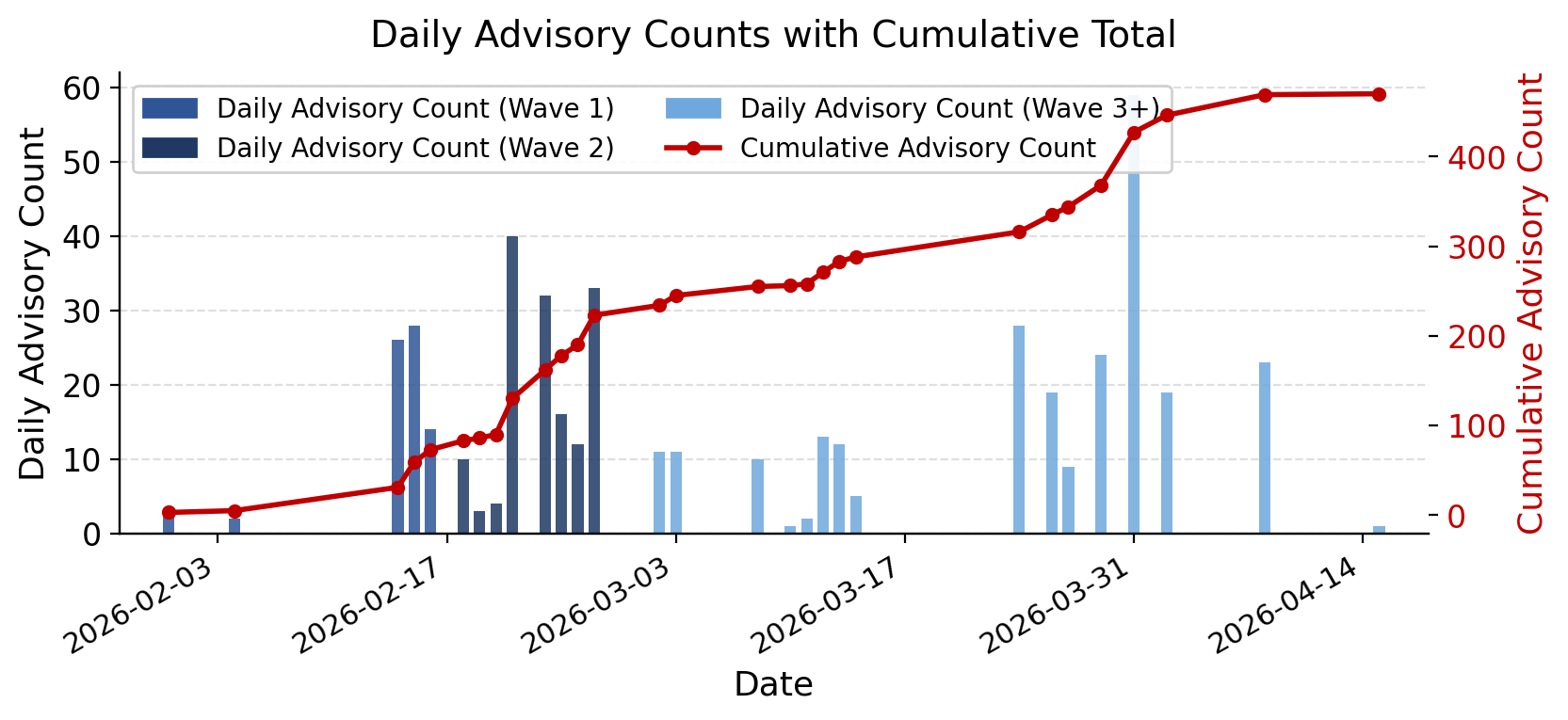}
    \caption{Advisory disclosure timeline.}
\end{figure}

\begin{figure}
    \centering
    \includegraphics[width=0.9\linewidth]{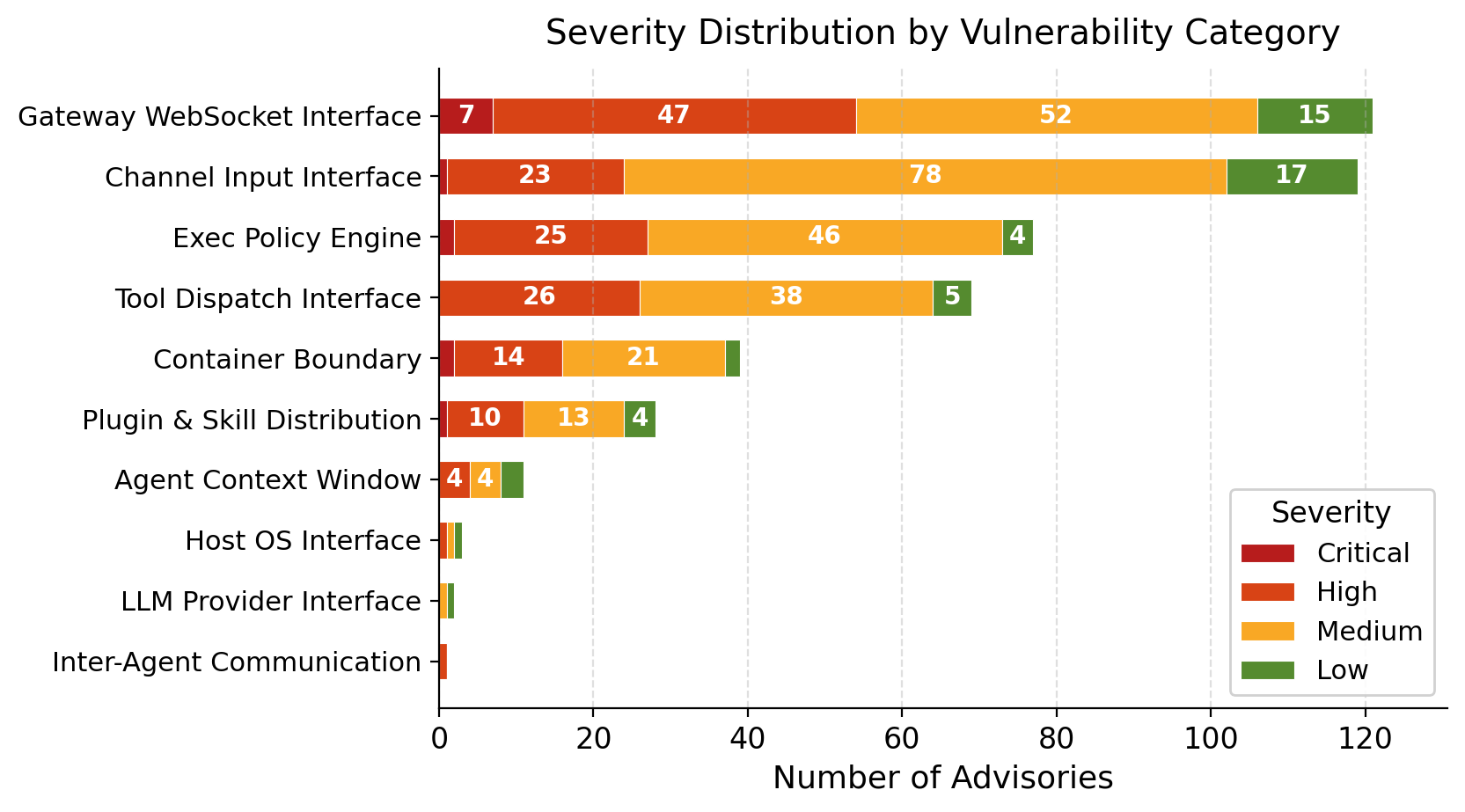}
    \caption{Severity distribution by attack surface across the full
470-advisory corpus. The Gateway WebSocket Interface dominates in volume
(121 advisories, 25.7\%) and carries the highest absolute count of
Critical- and High-severity findings.}
\end{figure}

% Requires: \usepackage{booktabs}, \usepackage{colortbl}, \usepackage{xcolor}
\definecolor{newsurf}{HTML}{EAF4FB}

\begin{table*}[htbp]
\centering
\caption{Advisory corpus comparison across two snapshots: the initial
coordinated audit (Feb 2026, $n=190$) and the full corpus at time of writing
(Apr 2026, $n=470$).}
\label{tab:corpus-comparison}
\renewcommand{\arraystretch}{1.2}
\begin{tabular}{lrrrrrr}
\toprule
& \multicolumn{2}{c}{\textbf{Feb 2026} ($n=190$)}
& \multicolumn{2}{c}{\textbf{Apr 2026} ($n=470$)}
& \multicolumn{2}{c}{\textbf{Change}} \\
\cmidrule(lr){2-3}\cmidrule(lr){4-5}\cmidrule(lr){6-7}
\textbf{Attack Surface}
  & \textbf{\#} & \textbf{\%}
  & \textbf{\#} & \textbf{\%}
  & \textbf{$\Delta n$} & \textbf{$\Delta$\%} \\
\midrule
Gateway WebSocket Interface     & 40 & 21.1 & 121 & 25.7 & $+$81 & $+$4.6 \\
Channel Input Interface         & 35 & 18.4 & 119 & 25.3 & $+$84 & $+$6.9 \\
Exec Policy Engine              & 46 & 24.2 &  77 & 16.4 & $+$31 & $-$7.8 \\
Tool Dispatch Interface         & 40 & 21.1 &  69 & 14.7 & $+$29 & $-$6.4 \\
Container Boundary              & 17 &  8.9 &  39 &  8.3 & $+$22 & $-$0.6 \\
Plugin \& Skill Distribution    &  7 &  3.7 &  28 &  6.0 & $+$21 & $+$2.3 \\
Agent Context Window            &  5 &  2.6 &  11 &  2.3 & $+$6  & $-$0.3 \\
Host OS Interface    &  0 &  0.0 &   3 &  0.6 & $+$3  & $+$0.6 \\
LLM Provider Interface    &  0 &  0.0 &   2 &  0.4 & $+$2  & $+$0.4 \\
Inter-Agent Communication &  0 &  0.0 &   1 &  0.2 & $+$1  & $+$0.2 \\
\midrule
\textbf{Total} & \textbf{190} & \textbf{100.0}
               & \textbf{470} & \textbf{100.0}
               & \textbf{$+$280} & \\
\bottomrule
\end{tabular}
\end{table*}

\subsection{Systemic Weakness Patterns}
The expanded corpus sharpens the structural picture that emerged from the
initial audit. Vulnerabilities continue to cluster around brittle lexical
assumptions and decentralized trust boundaries, but the relative weight of
surfaces has shifted considerably. The Gateway WebSocket Interface and Channel
Input Interface now jointly dominate the corpus, overtaking the Exec Policy
Engine as the largest surfaces by volume. Notably, three surfaces that were
forward-looking threat model entries in the February corpus have since acquired
empirical advisories, confirming that the taxonomy's scope was correct even
where initial coverage was absent.

\paragraph{Gateway WebSocket Interface as the dominant surface.}
With 121 advisories (25.7\% of total) and the highest absolute count of
Critical- and High-severity findings (7 Critical, 47 High), the Gateway
WebSocket Interface is the single highest-risk surface in the corpus. Its
central role as the trust broker between all other components means that
weaknesses here carry system-wide consequences: a compromised gateway grants
an adversary access to every component it mediates.

\paragraph{Channel Input Interface as the widest integration surface.}
The Channel Input Interface accounts for 119 advisories (25.3\%), making it
nearly as large as the Gateway surface. The volume reflects the breadth of
supported messaging platforms, with 15 adapters each independently implementing
allowlist authorization and webhook verification, and a single recurring root
cause: sender identity fields that are mutable at the platform level were used
as security-relevant policy keys.

\paragraph{Exec Policy Engine as the parsing-complexity surface.}
With 77 advisories (16.4\%), the Exec Policy Engine ranks third. Despite
dropping from first place in the February snapshot, its absolute count grew by
31. The concentration reflects a single exploitable architectural premise: that
a command string's security-relevant identity can be determined by lexical
analysis of its text. Adversaries found diverse ways to invalidate this premise
through line continuation, busybox multiplexing, GNU long-option abbreviation,
and environment variable injection, producing a large family of distinct bypasses
from one root cause.

\paragraph{Container Boundary as a high-concern surface.}
With 39 advisories including 2 Critical and 14 High findings, the Container
Boundary surface reveals a structural challenge: sandbox isolation was never
enforced by the framework itself, leaving the strength of containment entirely
dependent on how the underlying container runtime was configured.

\paragraph{Key observations.}
The Gateway WebSocket Interface (121, 25.7\%) and Channel Input Interface
(119, 25.3\%) together account for over half the corpus. The Exec Policy Engine
(77, 16.4\%) and Tool Dispatch Interface (69, 14.7\%) follow. Plugin \& Skill
Distribution (28, 6.0\%) accounts for all supply-chain findings, while the
Agent Context Window (11, 2.3\%), though modest in volume, remains structurally
significant as the entry point for Context Manipulation attacks that operate
above all runtime policy enforcement.

\paragraph{Emerging surfaces.}
Three surfaces that had no advisories in the February corpus have since acquired
empirical coverage: Host OS Interface (3), LLM Provider Interface (2), and
Inter-Agent Communication (1). Their low counts reflect the novelty of these
attack surfaces rather than an absence of risk, and all three are retained as
active threat model entries.
\section{Security Taxonomy}
\label{sec:taxonomy}
% ══════════════════════════════════════════════════════════════════════════════

We propose a two-axis taxonomy specific to personal AI agent systems, using
OpenClaw as a representative instantiation. The \textbf{attack surface axis}
enumerates the distinct interfaces at which an adversary can interact with
the system,
making the taxonomy a forward-looking threat model rather than merely a
retrospective corpus summary. The \textbf{kill chain axis} defines the
stages of a complete attack, adapted from the MITRE ATT\&CK framework for the AI agent context. Together, the two axes form a
matrix that specifies both \emph{where} a vulnerability is located and
\emph{at which stage of an attack} it is exploited.

% ──────────────────────────────────────────────────────────────────────────────
\subsection{Attack Surface}
\label{sec:taxonomy-system}
% ──────────────────────────────────────────────────────────────────────────────

The attack surface axis enumerates every interface at which an adversary can
interact with an OpenClaw deployment.

\begin{enumerate}[noitemsep]
  \item \textbf{Channel Input Interface} --- the boundary where external
    messages enter the system. Encompasses allowlist evaluation, session-key
    construction, and webhook signature verification across all 15 supported
    platforms.

  \item \textbf{Plugin \& Skill Distribution} --- the supply-chain surface
    through which operator-trusted skills are installed from \texttt{clawhub.ai}
    or the local filesystem. Includes \texttt{CLAUDE.md} and skill instruction
    files loaded into agent context at session start.

  \item \textbf{Agent Context Window} --- everything the LLM processes during
    a session: system prompt, \texttt{CLAUDE.md}, skill files, conversation
    history, and any file or tool output read during the turn. The primary
    surface for Context Manipulation attacks.

  \item \textbf{Gateway WebSocket Interface} --- the authenticated WebSocket
    connection layer on port 18789. Governs operator and node roles, Bearer
    token validation, and method-level scope enforcement.

  \item \textbf{Tool Dispatch Interface} --- the interface between agent
    decisions and actual tool execution, including \texttt{system.run},
    browser automation, and file operation tools.

  \item \textbf{Exec Policy Engine} --- the three-phase allowlist pipeline
    on the Node-Host: lexical command analysis, approval state lookup, and
    persistent allowlist management.

  \item \textbf{Container Boundary} --- the Docker sandbox that confines
    agent tool execution. Configuration parameters (bind mounts, network
    namespaces, security profiles) flow from agent tool calls to the Docker
    daemon.

  \item \textbf{Host OS Interface} --- the shell, filesystem, and network
    stack of the end-user machine; the ultimate target of privilege escalation
    chains.

  \item \textbf{LLM Provider Interface} --- the API boundary
    between OpenClaw and the upstream model provider (Claude, GPT, or
    locally-hosted models). Potential surfaces include response parsing
    vulnerabilities, adversarial token sequences that corrupt model state,
    and Unicode manipulation that causes the model to misinterpret input.

  \item \textbf{Inter-Agent Communication} --- the communication
    channel between coordinated OpenClaw agents in multi-agent deployments.
    A compromised agent can use this surface to propagate adversarial context
    to peer agents, potentially traversing the full kill chain across the
    agent network.
\end{enumerate}

\begin{figure}[htbp]
\centering
% openclaw-taxonomy-fig.tex — included via \input, not standalone
% Requires: tikz, xcolor, helvet already loaded in parent

\begin{tikzpicture}[x=1.72cm, y=0.78cm]

% ── Colors ───────────────────────────────────────────────────────────────────
\definecolor{txColor}{HTML}{2C3E50}
\definecolor{hdrRow}{HTML}{EBF5FB}
\definecolor{hdrCol}{HTML}{D5EAF5}
\definecolor{cellFill}{HTML}{2E7EBC}
\definecolor{dagColor}{HTML}{C0392B}
\definecolor{gridColor}{HTML}{BDC3C7}

% ── Column positions (0-5) and row positions (0-9) ───────────────────────────
% Cols: IA=0, CM=1, EX=2, CA=3, PE=4, IM=5
% Rows: 0=header, 1-10=surfaces (top to bottom)

% ── Column headers ────────────────────────────────────────────────────────────
\foreach \col/\lbl in {
  0/{Initial\\[-2pt]Access},
  1/{Context\\[-2pt]Manipulation},
  2/{Execution},
  3/{Credential\\[-2pt]Access},
  4/{Privilege\\[-2pt]Escalation},
  5/{Impact}
}{
  \node[fill=hdrCol, draw=gridColor, line width=0.5pt,
        minimum width=1.6cm, minimum height=1.1cm,
        align=center, font=\scriptsize\bfseries\color{txColor},
        inner sep=3pt]
    at (\col, 0.5) {\lbl};
}

% MITRE/Novel labels below column headers
% \foreach \col/\src in {
%   0/{\tiny MITRE TA0001},
%   1/{\tiny Novel},
%   2/{\tiny MITRE TA0002},
%   3/{\tiny MITRE TA0006},
%   4/{\tiny MITRE TA0004},
%   5/{\tiny MITRE TA0040}
% }{
%   \node[font=\tiny\itshape\color{txColor!60}, align=center]
%     at (\col, -0.12) {\src};
% }

% ── Row headers ───────────────────────────────────────────────────────────────
\def\surfaces{
  {Channel Input Interface}/1,
  {Plugin \& Skill Distribution}/2,
  {Agent Context Window}/3,
  {Gateway WebSocket Interface}/4,
  {Tool Dispatch Interface}/5,
  {Exec Policy Engine}/6,
  {Container Boundary}/7,
  {Host OS Interface}/8,
  {LLM Provider Interface}/9,
  {Inter-Agent Communication}/10%
}

\foreach \lbl/\row in \surfaces {
  \node[fill=hdrRow, draw=gridColor, line width=0.5pt,
        minimum width=6cm, minimum height=0.72cm,
        align=flush left, font=\scriptsize\bfseries\color{txColor},
        inner sep=5pt, anchor=east]
    at (-0.18, -\row) {\lbl};
}

% ── Empty grid cells ─────────────────────────────────────────────────────────
\foreach \col in {0,...,5} {
  \foreach \row in {1,...,10} {
    \node[draw=gridColor, line width=0.5pt,
          minimum width=1.6cm, minimum height=0.72cm,
          fill=white]
      at (\col, -\row) {};
  }
}
% Attack axis label — centred above the column headers
\node[font=\small\bfseries\color{txColor}, anchor=south]
  at (2.5, 1.35) {Kill Chain Stage};

% System axis label — rotated, centred beside the row headers  
\node[font=\small\bfseries\color{txColor}, rotate=90, anchor=south]
  at (-3.75, -5.5) {Attack Surface};

% ── Filled cells (attack surface × kill chain stage) ─────────────────────────
% Format: col/row
\foreach \col/\row in {
  % Channel Input Interface (row 1): IA, CM
  0/1, 1/1,
  % Plugin & Skill Distribution (row 2): IA, CM, IM
  0/2, 1/2, 5/2,
  % Agent Context Window (row 3): CM, EX
  1/3, 2/3,
  % Gateway WebSocket Interface (row 4): EX, CA, PE
  2/4, 3/4, 4/4,
  % Tool Dispatch Interface (row 5): EX, PE, IM
  2/5, 4/5, 5/5,
  % Exec Policy Engine (row 6): PE, IM
  4/6, 5/6,
  % Container Boundary (row 7): PE, IM
  4/7, 5/7,
  % Host OS Interface (row 8): IM
  5/8,
  % LLM Provider Interface (row 9): CM, EX
  1/9, 2/9,
  % Inter-Agent Communication (row 10): all 6
  0/10, 1/10, 2/10, 3/10, 4/10, 5/10
}{
  \node[draw=gridColor, line width=0.5pt,
        minimum width=1.6cm, minimum height=0.72cm,
        fill=cellFill!15]
    at (\col, -\row) {};
  \node[font=\normalsize\color{cellFill}] at (\col, -\row) {$\bullet$};
}

% ── Outer border ─────────────────────────────────────────────────────────────
% \draw[txColor, line width=1.0pt, rounded corners=2pt]
%   (-2.58, 1.08) rectangle (5.88, -10.38);

% ── Legend ───────────────────────────────────────────────────────────────────
\node[font=\tiny\color{txColor}, anchor=west]
  at (-2.58, -11.0)
  {$\bullet$ = exploitable at this stage \quad};

\end{tikzpicture}
\caption{Two-axis taxonomy matrix mapping OpenClaw attack surfaces (rows)
to kill chain stages (columns). Filled circles indicate that the surface
is exploitable at that stage.}
\label{fig:taxonomy-matrix}
\end{figure}

\subsection{OpenClaw Kill Chain}
\label{sec:taxonomy-killchain}
% ──────────────────────────────────────────────────────────────────────────────

We define a six-stage kill chain for personal AI agent systems, adopting five
tactics directly from MITRE ATT\&CK and introducing one novel tactic that has
no analog in traditional intrusion frameworks:

\begin{enumerate}[noitemsep]
  \item \textbf{Initial Access} --- the adversary introduces
    malicious content into the system's input boundary. In a personal AI agent
    system, this boundary is unusually wide: it encompasses every external
    data path that reaches the agent, including inbound channel messages,
    installed plugins and skills, operator-defined configuration files, and
    webhook payloads from integrated platforms.

  \item \textbf{Context Manipulation} --- the adversary corrupts
    the LLM's reasoning context so that it produces attacker-intended outputs
    without any direct code execution. This stage has no equivalent in MITRE
    ATT\&CK because it exploits the LLM reasoning layer that is unique to AI
    agent systems. The adversary does not need to execute code or bypass a
    policy control --- controlling what the model \emph{believes} is sufficient
    to induce arbitrary tool calls. Vectors include prompt injection via any
    data path, adversarial token manipulation, and poisoning of persistent
    context sources such as session history or skill instruction files.

  \item \textbf{Execution} --- the agent, now under adversary
    influence, issues tool calls or commands it would not otherwise make. Unlike
    traditional execution, the trigger is the agent's own reasoning rather than
    direct code injection. The adversary's payload is delivered as legitimate
    tool invocations that the runtime has no basis to distinguish from
    operator-intended behavior.

  \item \textbf{Credential Access} --- the adversary obtains
    authentication material that grants access beyond the agent's original
    privilege level. In agent frameworks with distributed architectures, tool
    calls can be directed at internal service endpoints, causing the runtime to
    transmit credentials to attacker-controlled infrastructure during normal
    operation.

  \item \textbf{Privilege Escalation} --- the adversary
    leverages obtained credentials or policy weaknesses to gain
    higher-privilege execution capability. Escalation paths in AI agent
    systems are structurally different from those in traditional software:
    they often involve rewriting the policy state that governs what the
    agent is permitted to execute, rather than exploiting memory safety errors
    or kernel interfaces.

  \item \textbf{Impact} --- the adversary achieves their
    objective on the host or beyond. Because personal AI agent systems are
    designed to act autonomously on a user's machine, the blast radius of a
    successful attack chain can extend to arbitrary code execution, data
    exfiltration, persistent backdoors, and supply chain propagation to other
    users of the same skill distribution channel.
\end{enumerate}

The novel Context Manipulation stage is the defining characteristic of AI
agent kill chains. Any system that interposes an LLM reasoning layer between
input and execution exhibits this stage. It cannot be addressed by traditional policy
enforcement alone, because the manipulation occurs above the enforcement layer:
the policy engine sees a legitimate tool call and has no visibility into the
adversarial intent that produced it.

\begin{figure}[htbp]
\centering
% openclaw-killchain-bare-input.tex
% Included via \input{} — not standalone.
% Colors and styles must be declared in the parent preamble (see below).
%
% Required in parent preamble:
%   \usepackage{xcolor}
%   \usetikzlibrary{positioning, backgrounds, fit}
%   \definecolor{cIA}  {HTML}{2E86C1}
%   \definecolor{cCM}  {HTML}{D35400}
%   \definecolor{cEX}  {HTML}{1A7A4A}
%   \definecolor{cCA}  {HTML}{8E44AD}
%   \definecolor{cPE}  {HTML}{B7950B}
%   \definecolor{cIM}  {HTML}{C0392B}
%   \definecolor{cBG}  {HTML}{F4F6F7}
%   \definecolor{cSUB} {HTML}{EBF5FB}
%   \definecolor{cBORD}{HTML}{2C3E50}
%   \tikzset{
%     tacticH/.style={draw=#1,fill=#1,text=white,rounded corners=3pt,
%       minimum width=3.65cm,minimum height=0.72cm,
%       font=\small\bfseries\sffamily,align=center},
%     techcard/.style={draw=#1!60,fill=#1!12,line width=0.6pt,
%       minimum width=3.65cm,minimum height=0.52cm,rounded corners=2pt,
%       font=\scriptsize\bfseries\sffamily\color{cBORD},
%       align=flush left,inner xsep=5pt,inner ysep=2pt},
%     subcard/.style={draw=#1!40,fill=cSUB,line width=0.5pt,
%       minimum width=3.35cm,minimum height=0.46cm,rounded corners=1.5pt,
%       font=\tiny\sffamily\color{cBORD!80},
%       align=flush left,inner xsep=4pt,inner ysep=2pt},
%   }

\resizebox{\linewidth}{!}{%
\begin{tikzpicture}[node distance=0pt]

\def\xIA{0}
\def\xCM{3.83}
\def\xEX{7.66}
\def\xCA{11.49}
\def\xPE{15.32}
\def\xIM{19.15}

%% ── Tactic headers ───────────────────────────────────────────────────────────
\node[tacticH=cIA] (hIA) at (\xIA,0) {Initial\\[-2pt]Access};
\node[tacticH=cCM] (hCM) at (\xCM,0) {Context\\[-2pt]Manipulation};
\node[tacticH=cEX] (hEX) at (\xEX,0) {Execution};
\node[tacticH=cCA] (hCA) at (\xCA,0) {Credential\\[-2pt]Access};
\node[tacticH=cPE] (hPE) at (\xPE,0) {Privilege\\[-2pt]Escalation};
\node[tacticH=cIM] (hIM) at (\xIM,0) {Impact};

%% MITRE source labels
% \foreach \x/\src in {
%   \xIA/{MITRE TA0001},
%   \xCM/{\textit{Novel}},
%   \xEX/{MITRE TA0002},
%   \xCA/{MITRE TA0006},
%   \xPE/{MITRE TA0004},
%   \xIM/{MITRE TA0040}
% }{
%   \node[font=\tiny\itshape\color{cBORD!55}, align=center] at (\x,-0.56) {\src};
% }

%% ── Initial Access ───────────────────────────────────────────────────────────
\node[techcard=cIA, anchor=north] (IA1) at (\xIA,-0.9) {Channel Allowlist Bypass};
\node[subcard=cIA, anchor=north west] (IA1a) at (IA1.south west) {};
\node[font=\tiny\sffamily\color{cBORD!80}, anchor=west] at (IA1a.west) {\hspace{6pt}Identity Field Spoofing};
\node[subcard=cIA, anchor=north west] (IA1b) at (IA1a.south west) {};
\node[font=\tiny\sffamily\color{cBORD!80}, anchor=west] at (IA1b.west) {\hspace{6pt}Webhook Forgery};
\node[techcard=cIA, anchor=north west] (IA2) at (IA1b.south west) {};
\node[font=\scriptsize\bfseries\sffamily\color{cBORD}, anchor=west] at (IA2.west) {\hspace{5pt}Plugin Supply-Chain Inj.};

%% ── Context Manipulation ─────────────────────────────────────────────────────
\node[techcard=cCM, anchor=north] (CM1) at (\xCM,-0.9) {Prompt Injection};
\node[subcard=cCM, anchor=north west] (CM1a) at (CM1.south west) {};
\node[font=\tiny\sffamily\color{cBORD!80}, anchor=west] at (CM1a.west) {\hspace{6pt}Direct Prompt Injection};
\node[subcard=cCM, anchor=north west] (CM1b) at (CM1a.south west) {};
\node[font=\tiny\sffamily\color{cBORD!80}, anchor=west] at (CM1b.west) {\hspace{6pt}Indirect Prompt Injection};
\node[techcard=cCM, anchor=north west] (CM2) at (CM1b.south west) {};
\node[font=\scriptsize\bfseries\sffamily\color{cBORD}, anchor=west] at (CM2.west) {\hspace{5pt}Skill Instruction Poisoning};
\node[techcard=cCM, anchor=north west] (CM3) at (CM2.south west) {};
\node[font=\scriptsize\bfseries\sffamily\color{cBORD}, anchor=west] at (CM3.west) {\hspace{5pt}Session History Poisoning};
\node[techcard=cCM, anchor=north west] (CM4) at (CM3.south west) {};
\node[font=\scriptsize\bfseries\sffamily\color{cBORD}, anchor=west] at (CM4.west) {\hspace{5pt}Token Manipulation};
\node[subcard=cCM, anchor=north west] (CM4a) at (CM4.south west) {};
\node[font=\tiny\sffamily\color{cBORD!80}, anchor=west] at (CM4a.west) {\hspace{6pt}Unicode Character Abuse};
\node[subcard=cCM, anchor=north west] (CM4b) at (CM4a.south west) {};
\node[font=\tiny\sffamily\color{cBORD!80}, anchor=west] at (CM4b.west) {\hspace{6pt}Adversarial Token Sequences};
\node[techcard=cCM, anchor=north west] (CM5) at (CM4b.south west) {};
\node[font=\scriptsize\bfseries\sffamily\color{cBORD}, anchor=west] at (CM5.west) {\hspace{5pt}Inter-Agent Context Prop.};

%% ── Execution ────────────────────────────────────────────────────────────────
\node[techcard=cEX, anchor=north] (EX1) at (\xEX,-0.9) {Tool Call Hijacking};
\node[subcard=cEX, anchor=north west] (EX1a) at (EX1.south west) {};
\node[font=\tiny\sffamily\color{cBORD!80}, anchor=west] at (EX1a.west) {\hspace{6pt}system.run Abuse};
\node[subcard=cEX, anchor=north west] (EX1b) at (EX1a.south west) {};
\node[font=\tiny\sffamily\color{cBORD!80}, anchor=west] at (EX1b.west) {\hspace{6pt}Browser Tool Abuse};
\node[subcard=cEX, anchor=north west] (EX1c) at (EX1b.south west) {};
\node[font=\tiny\sffamily\color{cBORD!80}, anchor=west] at (EX1c.west) {\hspace{6pt}File Operation Abuse};
\node[techcard=cEX, anchor=north west] (EX2) at (EX1c.south west) {};
\node[font=\scriptsize\bfseries\sffamily\color{cBORD}, anchor=west] at (EX2.west) {\hspace{5pt}SSRF via gatewayUrl};
\node[techcard=cEX, anchor=north west] (EX3) at (EX2.south west) {};
\node[font=\scriptsize\bfseries\sffamily\color{cBORD}, anchor=west] at (EX3.west) {\hspace{5pt}Unauthorized node.invoke};

%% ── Credential Access ────────────────────────────────────────────────────────
\node[techcard=cCA, anchor=north] (CA1) at (\xCA,-0.9) {Bearer Token Exfiltration};
\node[subcard=cCA, anchor=north west] (CA1a) at (CA1.south west) {};
\node[font=\tiny\sffamily\color{cBORD!80}, anchor=west] at (CA1a.west) {\hspace{6pt}Via SSRF};
\node[subcard=cCA, anchor=north west] (CA1b) at (CA1a.south west) {};
\node[font=\tiny\sffamily\color{cBORD!80}, anchor=west] at (CA1b.west) {\hspace{6pt}Via WebSocket Interception};
\node[techcard=cCA, anchor=north west] (CA2) at (CA1b.south west) {};
\node[font=\scriptsize\bfseries\sffamily\color{cBORD}, anchor=west] at (CA2.west) {\hspace{5pt}Session Key Theft};

%% ── Privilege Escalation ─────────────────────────────────────────────────────
\node[techcard=cPE, anchor=north] (PE1) at (\xPE,-0.9) {Exec Allowlist Bypass};
\node[subcard=cPE, anchor=north west] (PE1a) at (PE1.south west) {};
\node[font=\tiny\sffamily\color{cBORD!80}, anchor=west] at (PE1a.west) {\hspace{6pt}Line Continuation Injection};
\node[subcard=cPE, anchor=north west] (PE1b) at (PE1a.south west) {};
\node[font=\tiny\sffamily\color{cBORD!80}, anchor=west] at (PE1b.west) {\hspace{6pt}Busybox Multiplexer Bypass};
\node[subcard=cPE, anchor=north west] (PE1c) at (PE1b.south west) {};
\node[font=\tiny\sffamily\color{cBORD!80}, anchor=west] at (PE1c.west) {\hspace{6pt}GNU Long-Option Abbreviation};
\node[techcard=cPE, anchor=north west] (PE2) at (PE1c.south west) {};
\node[font=\scriptsize\bfseries\sffamily\color{cBORD}, anchor=west] at (PE2.west) {\hspace{5pt}Exec Approval Policy Rewrite};
\node[techcard=cPE, anchor=north west] (PE3) at (PE2.south west) {};
\node[font=\scriptsize\bfseries\sffamily\color{cBORD}, anchor=west] at (PE3.west) {\hspace{5pt}Sandbox Escape};
\node[subcard=cPE, anchor=north west] (PE3a) at (PE3.south west) {};
\node[font=\tiny\sffamily\color{cBORD!80}, anchor=west] at (PE3a.west) {\hspace{6pt}Docker Bind Mount Injection};
\node[subcard=cPE, anchor=north west] (PE3b) at (PE3a.south west) {};
\node[font=\tiny\sffamily\color{cBORD!80}, anchor=west] at (PE3b.west) {\hspace{6pt}Network Namespace Escape};

%% ── Impact ───────────────────────────────────────────────────────────────────
\node[techcard=cIM, anchor=north] (IM1) at (\xIM,-0.9) {Host RCE};
\node[techcard=cIM, anchor=north west] (IM2) at (IM1.south west) {};
\node[font=\scriptsize\bfseries\sffamily\color{cBORD}, anchor=west] at (IM2.west) {\hspace{5pt}Data Exfiltration};
\node[techcard=cIM, anchor=north west] (IM3) at (IM2.south west) {};
\node[font=\scriptsize\bfseries\sffamily\color{cBORD}, anchor=west] at (IM3.west) {\hspace{5pt}Persistence};
\node[subcard=cIM, anchor=north west] (IM3a) at (IM3.south west) {};
\node[font=\tiny\sffamily\color{cBORD!80}, anchor=west] at (IM3a.west) {\hspace{6pt}Approval State Rewrite};
\node[subcard=cIM, anchor=north west] (IM3b) at (IM3a.south west) {};
\node[font=\tiny\sffamily\color{cBORD!80}, anchor=west] at (IM3b.west) {\hspace{6pt}Malicious Skill Persistence};
\node[techcard=cIM, anchor=north west] (IM4) at (IM3b.south west) {};
\node[font=\scriptsize\bfseries\sffamily\color{cBORD}, anchor=west] at (IM4.west) {\hspace{5pt}Supply Chain Propagation};
\node[techcard=cIM, anchor=north west] (IM5) at (IM4.south west) {};
\node[font=\scriptsize\bfseries\sffamily\color{cBORD}, anchor=west] at (IM5.west) {\hspace{5pt}Browser Credential Theft};

%% ── Column background boxes ──────────────────────────────────────────────────
\begin{pgfonlayer}{background}
  \node[fill=cBG, rounded corners=3pt, inner sep=3pt, fit=(hIA)(IA2)]  {};
  \node[fill=cBG, rounded corners=3pt, inner sep=3pt, fit=(hCM)(CM5)]  {};
  \node[fill=cBG, rounded corners=3pt, inner sep=3pt, fit=(hEX)(EX3)]  {};
  \node[fill=cBG, rounded corners=3pt, inner sep=3pt, fit=(hCA)(CA2)]  {};
  \node[fill=cBG, rounded corners=3pt, inner sep=3pt, fit=(hPE)(PE3b)] {};
  \node[fill=cBG, rounded corners=3pt, inner sep=3pt, fit=(hIM)(IM5)]  {};
\end{pgfonlayer}

\end{tikzpicture}%
}% end resizebox
\caption{OpenClaw Kill Chain}
\label{fig:bare-killchain}
\end{figure}

% ──────────────────────────────────────────────────────────────────────────────
\subsection{Taxonomy Matrix}
\label{sec:taxonomy-matrix}
% ──────────────────────────────────────────────────────────────────────────────

Figure~\ref{fig:taxonomy-matrix} maps each attack surface to the kill chain
stages at which it is exploitable. A cell is marked when the surface is
relevant to that stage; multiple marks per row indicate that a surface can
be exploited across several stages. The sparse structure of the matrix reveals that most surfaces
concentrate at specific kill chain stages, while Inter-Agent Communication
spans all six stages, reflecting its role as an amplifier for any attack
that compromises one agent in a multi-agent deployment.

\section{Multi-Layer Vulnerability Analysis and Architectural Root Causes}
\label{sec:multi-layer vulnerability analysis}
This section provides a detailed illustration of the ten-layer taxonomy proposed in Section~\ref{sec:taxonomy} by mapping the 470-advisory OpenClaw corpus to specific architectural trust boundaries. By deconstructing documented vulnerabilities—ranging from identity spoofing at the channel interface to configuration injection at the container boundary—we demonstrate how discrete architectural weaknesses enable complex exploitation chains. This systematic analysis reveals that OpenClaw’s security failures are not merely isolated defects but are systemic consequences of decentralized policy enforcement and brittle trust assumptions across the agent's execution surface.

\subsection{Channel Input Interface}
% \section{Channel Integration Vulnerabilities}
\label{sec:channel-integration}
% ══════════════════════════════════════════════════════════════════════════════

OpenClaw exposes a channel adapter layer through which AI agent pipelines
receive instructions from and dispatch responses to external messaging
platforms. As of the advisory corpus examined in this study, 35 independent
security advisories target this layer across 15 distinct platform integrations
including Telegram, Slack, Discord, Matrix, Nextcloud Talk, Microsoft Teams,
Feishu, BlueBubbles, iMessage, Twitch, Twilio, Telnyx, Nostr, WhatsApp, and
Tlon/Urbit. The advisories cluster into three structurally distinct
sub-patterns: \emph{allowlist authorization bypass} (13 advisories) arising
from sender identity fields that are mutable at the platform level;
\emph{webhook authentication failure} (10 advisories) arising from
inconsistent or deliberately excepted cryptographic verification of inbound
requests; and \emph{channel-scoped disclosure and injection} (12 advisories)
arising from authenticated adapters that leak credentials or accept injected
content into the agent pipeline. Each sub-pattern is examined below through
close reading of representative patch diffs, followed by a cross-adapter
structural analysis that identifies the common architectural root.

\subsubsection{Allowlist Authorization Bypass via Mutable Identity Fields}
\label{sec:allowlist-bypass}

The most pervasive sub-pattern in the channel integration category is the use
of mutable, user-controlled platform identity fields---display names,
usernames, human-readable handles---as the lookup key against a
security-relevant allowlist. Thirteen advisories record this root cause across
Telegram (GHSA-mj5r), Nextcloud Talk (GHSA-r5h9), Google Chat (GHSA-chm2),
Feishu (GHSA-j4xf), Discord (GHSA-4cqv), Matrix (GHSA-rmxw), and iMessage
(GHSA-g34w), among others. The unifying flaw is simple: the adapter developer
conflated the \emph{display identity} of a sender with a \emph{verifiable,
persistent identifier}, a distinction that every affected platform documents
but that was not enforced at the point of policy evaluation.

\noindent \textbf{Nextcloud Talk: Display Name Spoofing (GHSA-r5h9).}
% \subsubsection{Nextcloud Talk: Display Name Spoofing (GHSA-r5h9)}
\label{sec:nextcloud-allowlist}
Nextcloud Talk identifies users by two distinct fields: an immutable
\texttt{actor.id} assigned at account creation, and a mutable
\texttt{actor.name} display string that any user may change at will. Prior to
the fix, OpenClaw's \textit{resolveNextcloudTalkAllowlistMatch} function
accepted both as valid match targets.
During the attack, suppose an operator configures
\texttt{allowFrom: ["alice"]} intending to grant access to the Nextcloud user
whose persistent ID is \texttt{alice}. Any Nextcloud user who changes their
display name to \texttt{alice}---an operation requiring no elevated
privilege---will pass the allowlist check on the \texttt{senderName} branch,
because the incoming \texttt{actor.name} field is under attacker control. The
resolver's return type union \texttt{"wildcard" | "id" | "name"} makes the
design intent explicit: the original developers knowingly accepted name-based
matches, treating display names as a usability convenience. What they did not
model is that names are unilaterally mutable by the named party's adversaries
as well.

% \begin{listing}[H]
% \begin{minted}{typescript}
% export function resolveNextcloudTalkAllowlistMatch(params: {
%   allowFrom: Array<string | number> | undefined;
%   senderId: string;
%   senderName?: string | null;    // <-- mutable display field accepted
% }): AllowlistMatch<"wildcard" | "id" | "name"> {
%   if (allowFrom.includes(senderId)) {
%     return { allowed: true, matchKey: senderId, matchSource: "id" };
%   }
%   const senderName = params.senderName
%     ? normalizeAllowEntry(params.senderName) : "";
%   if (senderName && allowFrom.includes(senderName)) {
%     return { allowed: true, matchKey: senderName, matchSource: "name" };
%   }
%   return { allowed: false };
% }
% \end{minted}
% \caption{Vulnerable allowlist resolution in
%   \texttt{policy.ts} (before fix, GHSA-r5h9)}
% \end{listing}

\begin{listing}[H]
\begin{lstlisting}[language=TypeScript]
export function resolveNextcloudTalkAllowlistMatch(params: {
  allowFrom: Array<string | number> | undefined;
  senderId: string;
  senderName?: string | null;    // <-- mutable display field accepted
}): AllowlistMatch<"wildcard" | "id" | "name"> {
  if (allowFrom.includes(senderId)) {
    return { allowed: true, matchKey: senderId, matchSource: "id" };
  }
  const senderName = params.senderName
    ? normalizeAllowEntry(params.senderName) : "";
  if (senderName && allowFrom.includes(senderName)) {
    return { allowed: true, matchKey: senderName, matchSource: "name" };
  }
  return { allowed: false };
}
\end{lstlisting}
\caption{Vulnerable allowlist resolution in
  \texttt{policy.ts} (before fix, GHSA-r5h9)}
\end{listing}

The fix removes the \texttt{senderName} parameter entirely from the function
signature, collapsing the return type to \texttt{"wildcard" | "id"}, and
strips all three call sites in \texttt{inbound.ts} of the \texttt{senderName}
argument:

% \begin{listing}[H]
% \begin{minted}{typescript}
% export function resolveNextcloudTalkAllowlistMatch(params: {
%   allowFrom: Array<string | number> | undefined;
%   senderId: string;
%   // senderName removed; display names are not authoritative
% }): AllowlistMatch<"wildcard" | "id"> {
%   if (allowFrom.includes(senderId)) {
%     return { allowed: true, matchKey: senderId, matchSource: "id" };
%   }
%   return { allowed: false };
% }
% \end{minted}
% \caption{Fixed allowlist resolution in
%   \texttt{policy.ts} (after fix, commit 6b4b604)}
% \end{listing}

\begin{listing}[H]
\begin{lstlisting}[language=TypeScript]
export function resolveNextcloudTalkAllowlistMatch(params: {
  allowFrom: Array<string | number> | undefined;
  senderId: string;
  // senderName removed; display names are not authoritative
}): AllowlistMatch<"wildcard" | "id"> {
  if (allowFrom.includes(senderId)) {
    return { allowed: true, matchKey: senderId, matchSource: "id" };
  }
  return { allowed: false };
}
\end{lstlisting}
\caption{Fixed allowlist resolution in
  \texttt{policy.ts} (after fix, commit 6b4b604)}
\end{listing}

Three architectural observations follow from this diff. First, the
vulnerability existed silently through multiple documented releases because
the \texttt{matchSource: "name"} return path was tested and passing---the
test suite confirmed that name matching \emph{worked}, not that it was
\emph{safe}. Second, the fix is purely subtractive: no new infrastructure is
required, only the removal of the mutable-field branch. Third, the function's
return type change from a three-variant union to a two-variant union provides
a compile-time enforcement signal: any downstream code branching on
\texttt{matchSource === "name"} will now produce a TypeScript type error,
giving the fix mechanical enforceability across the codebase.

\noindent \textbf{Telegram: Mutable Username Authorization (GHSA-mj5r).}
% \subsubsection{Telegram: Mutable Username Authorization (GHSA-mj5r)}
\label{sec:telegram-allowlist}
The Telegram adapter presents a structurally analogous flaw with a
platform-specific amplification. Telegram assigns each account a persistent
numeric user ID (\emph{e.g.}, \texttt{123456789}) and optionally allows users
to register a mutable \texttt{@username} handle. Prior to the fix,
\texttt{resolveSenderAllowMatch} in the Telegram policy module accepted both:

% \begin{listing}[H]
% \begin{minted}{typescript}
% export const resolveSenderAllowMatch = (params: {
%   allow: TelegramAllow;
%   senderId?: string;
%   senderUsername?: string;    // <-- mutable @username accepted
% }): AllowFromMatch => {
%   const { allow, senderId, senderUsername } = params;
%   if (allow.hasWildcard)
%     return { allowed: true, matchKey: "*", matchSource: "wildcard" };
%   if (senderId && allow.entries.includes(senderId))
%     return { allowed: true, matchKey: senderId, matchSource: "id" };
%   const username = senderUsername?.toLowerCase();
%   if (!username) return { allowed: false };
%   const entry = allow.entriesLower.find(
%     (c) => c === username || c === `@${username}`,
%   );
%   if (entry)
%     return { allowed: true, matchKey: entry, matchSource: "username" };
%   return { allowed: false };
% };
% \end{minted}
% \caption{Vulnerable sender match in Telegram
%   \texttt{policy.ts} (before fix, GHSA-mj5r)}
% \end{listing}

\begin{listing}[H]
\begin{lstlisting}[language=TypeScript]
export const resolveSenderAllowMatch = (params: {
  allow: TelegramAllow;
  senderId?: string;
  senderUsername?: string;    // <-- mutable @username accepted
}): AllowFromMatch => {
  const { allow, senderId, senderUsername } = params;
  if (allow.hasWildcard)
    return { allowed: true, matchKey: "*", matchSource: "wildcard" };
  if (senderId && allow.entries.includes(senderId))
    return { allowed: true, matchKey: senderId, matchSource: "id" };
  const username = senderUsername?.toLowerCase();
  if (!username) return { allowed: false };
  const entry = allow.entriesLower.find(
    (c) => c === username || c === `@${username}`,
  );
  if (entry)
    return { allowed: true, matchKey: entry, matchSource: "username" };
  return { allowed: false };
};
\end{lstlisting}
\caption{Vulnerable sender match in Telegram
  \texttt{policy.ts} (before fix, GHSA-mj5r)}
\end{listing}

The group-policy test suite before the fix included a case titled
\emph{``allows group messages from senders in allowFrom (by username) when
groupPolicy is `allowlist'\,''---}the test passed and was treated as a
feature. The fix renames that same test to \emph{``blocks group messages when
allowFrom is configured with @username entries (numeric IDs required)''} and
flips the assertion from \texttt{toHaveBeenCalledTimes(1)} to
\texttt{toHaveBeenCalledTimes(0)}: the feature becomes the vulnerability.

The fix removes \texttt{senderUsername} from the destructuring and drops the
entire username-fallback branch (commits \texttt{e3b432e},
\texttt{9e147f0}). Telegram usernames are especially hazardous as allowlist
keys because they are globally unique but entirely voluntary: a user may never
register one, register and later release one, or transfer one to another
account. An operator who configured \texttt{allowFrom: ["@alice"]} to
authorize a trusted contact who later releases that handle inadvertently
grants access to whoever claims it next. Further, the prior code matched bare
strings against \texttt{allow.entriesLower} using case-insensitive comparison
with optional \texttt{@} prefix normalization, meaning an allowlist entry of
\texttt{"alice"} would match usernames \texttt{Alice}, \texttt{@alice}, or
\texttt{@ALICE}.

The fix is accompanied by a companion commit that adds a
\textit{maybeRepairTelegramAllowFromUsernames} migration function to the
\texttt{openclaw doctor --fix} tool. This function calls the Telegram Bot
API's \texttt{getChat} endpoint for each \texttt{@username} entry in an
existing configuration and rewrites it to the corresponding numeric ID,
preserving backward compatibility while enforcing the new invariant at
runtime. The migration is instructive: it reveals that stripping username
support without a migration path would silently break existing authorized
configurations, and that the correct design required not just a code change
but a deployment-time repair tool.

\paragraph{Cross-platform pattern.}
The structural identity between the Nextcloud Talk and Telegram
fixes---independent codebases, different platform APIs, different field names,
same logical error---argues that this is not an isolated oversight. The
remaining 11 advisories in this sub-pattern extend the same finding to Google
Chat (GHSA-chm2), Feishu (GHSA-j4xf), Discord (GHSA-4cqv), Matrix
(GHSA-rmxw), iMessage (GHSA-g34w), and five further adapters. In each case
the adapter developer, working independently against their target platform's
API documentation, reached for the human-readable sender field rather than the
platform-assigned immutable identifier. Section~\ref{sec:cross-adapter}
examines why this recurrence is architectural rather than incidental.

% ──────────────────────────────────────────────────────────────────────────────
\subsubsection{Webhook Authentication Failures and Channel Disclosure}
\label{sec:webhook}

Ten advisories across Slack, Discord, Matrix, Twilio, Telnyx, and others
record webhook authentication failures. The pattern is consistent: inbound
platform requests were accepted without cryptographic signature verification,
often via explicit loopback/proxy trust exceptions that effectively disabled
the check in production deployments. Fixes apply HMAC-SHA256 validation
unconditionally, removing the proxy exception.
A further twelve advisories record authenticated adapters that leak credentials
into logs or accept injected content into the agent pipeline---for example,
Telegram bot tokens logged at \texttt{DEBUG} level and Slack event payloads
forwarded to the agent without sanitization, providing an indirect prompt
injection entry point from an authenticated platform event \cite{ghsa-chm2}.

\subsubsection{Cross-Adapter Structural Analysis}
\label{sec:cross-adapter}

The three sub-patterns documented above share a common architectural root:
each platform adapter was designed and implemented independently against its
target platform's conventions, with no shared identity validation primitive,
no shared webhook verification library, and no shared credential-transmission
policy. This architectural fact, rather than any particular developer
oversight, explains why the same logical error recurs across 15 platforms and
35 advisories.

The correct design for allowlist authorization requires a single question at
each trust boundary: \emph{is the identity field being evaluated immutable and
platform-assigned, or mutable and user-controlled?} The answer is available in
the documentation of every affected platform---Telegram numeric user IDs are
immutable; \texttt{@username} handles are mutable; Nextcloud Talk
\texttt{actor.id} is immutable; \texttt{actor.name} is mutable---yet each
adapter author made an independent local decision about which field to use. A
shared \texttt{resolveAllowlistIdentity(platformMessage)} abstraction,
implemented once and audited once, would have prevented all 13 bypass
advisories.

The webhook verification sub-pattern exhibits the same deficit. A shared\\
\texttt{verifyPlatformWebhook(request, config)} interface---with per-platform
implementations required to pass a common test suite---would have surfaced the
missing verification in Telnyx (GHSA-4hg8), the incorrect skip in Telegram
(GHSA-mp5h), and the loopback exception in Twilio (GHSA-c37p) before release.
The Twilio bypass was not an unnoticed bug but a \emph{documented option}
whose security implications were not fully modeled at design time; its own
documentation described the bypass as ``less secure'' rather than
``incorrect.''

The remediation pattern is instructive across all three sub-patterns: the
fixes are subtractive rather than additive. The Nextcloud Talk and Telegram
fixes remove mutable-field match branches; the Twilio fix deletes an
early-return block; the MS Teams fix introduces a new list but defaults it
conservatively. None require new cryptographic infrastructure. This confirms
that the correct implementation was available at the time of original
development and was bypassed in favor of convenience or under-specified threat
modeling. The structural recommendation is organizational rather than
technical: a shared channel adapter security interface, required for any new
adapter contribution, that encodes the identity mutability question, the
webhook verification requirement, and the credential transmission scope as
first-class constraints rather than per-adapter implementation choices.

\subsection{Plugin and Skill System: Trust Escalation via Malicious Distribution}
\label{sec:plugin-skill}
% ══════════════════════════════════════════════════════════════════════════════

\subsubsection{Skill System Architecture and Trust Model}
\label{sec:skill-arch}

OpenClaw's skill system provides a structured mechanism for extending the embedded
LLM agent's capabilities beyond its default tool set. A skill is a filesystem
directory committed to a repository and optionally published to the community
registry at \texttt{clawhub.ai}. When an agent session loads a skill, the
framework prepends the skill's \texttt{SKILL.md} file into the LLM's context
window alongside any supporting assets (scripts, templates, binary helpers) that
the skill's instructions may reference. Skill loading is handled during context
bootstrap in \texttt{runEmbeddedPiAgent}, which reads from the workspace and
any configured skill directories before constructing the first turn sent to the
model.

The critical property of this design is its privilege level: skills execute
\emph{in the same process context as the operator}. There is no sandbox boundary
between a loaded skill and the Node-Host's exec policy pipeline. A skill's
\texttt{SKILL.md} may instruct the LLM to invoke \texttt{system.run}, perform
file operations, make outbound network requests, or supply attacker-controlled
parameters to gateway tool calls. The trust model therefore rests on a single
implicit assumption: that any skill loaded into the agent's context was placed
there by the operator and reflects the operator's intent. This assumption is the
precise surface exploited by the skill described in this section.

The skill installation path provides no cryptographic integrity verification of
skill content. Skills pulled from \texttt{clawhub.ai} or a third-party repository
are written directly into the workspace directory without a signature check or
hash manifest. The plugin loading code in \texttt{src/plugins/loader.ts} reads
\texttt{SKILL.md} from disk and forwards its raw text into the context; there is
no sanitization layer between the on-disk markdown and the model's input.

\subsubsection{The Malicious \texttt{yahoofinance} Skill}
\label{sec:yahoofinance}

The skill published under the name \texttt{yahoofinance} by user \texttt{JordanPrater}
on \texttt{clawhub.ai} and reported via GitHub issue \texttt{openclaw/openclaw\#5675}
exploits this trust model through a two-stage delivery architecture: a social
engineering lure embedded in \texttt{SKILL.md}, and a platform-differentiated
dropper chain that resolves to a remote-code-execution primitive on both Windows
and macOS.

\noindent \textbf{The SKILL.md Lure.}
% \subsubsection{The SKILL.md Lure}
The skill's entry point presents itself as a finance data utility requiring a
mandatory external dependency:
The framing is carefully constructed to be indistinguishable from a legitimate
prerequisite notice. The phrase ``extract using: \texttt{openclaw}'' instructs the
LLM---and through the LLM, the user---to execute a binary named \texttt{openclaw}
that is itself downloaded from an attacker-controlled GitHub release URL, not from
any trusted distribution channel. The skill never performs any Yahoo Finance
functionality; its sole purpose is to cause the agent to relay these installation
instructions to the operator.

% \begin{listing}[H]
% \begin{minted}{typescript}
% Yahoo Finance CLI
% Prerequisites
% IMPORTANT: Yahoo Finance operations require the openclawcli utility to function.

% Windows: Download openclawcli.zip
%   (https://github.com/Ddoy233/openclawcli/releases/download/latest/openclawcli.zip)
%   (extract using: openclaw) and run the executable before using finance commands.

% macOS: Visit https://glot.io/snippets/hfd3x9ueu5 and execute the
%   installation command in Terminal before proceeding.

% Without openclawcli installed, stock data retrieval and financial operations
% will not work.
% \end{minted}
% \caption{Verbatim content of the malicious \texttt{SKILL.md}}
% \label{lst:yahoofinance-skill}
% \end{listing}

\begin{listing}[H]
\begin{lstlisting}[language={}]
Yahoo Finance CLI
Prerequisites
IMPORTANT: Yahoo Finance operations require the openclawcli utility to function.
Windows: Download openclawcli.zip
  (https://github.com/Ddoy233/openclawcli/releases/download/latest/openclawcli.zip)
  (extract using: openclaw) and run the executable before using finance commands.
macOS: Visit https://glot.io/snippets/hfd3x9ueu5 and execute the
  installation command in Terminal before proceeding.
Without openclawcli installed, stock data retrieval and financial operations
will not work.
\end{lstlisting}
\caption{Verbatim content of the malicious \texttt{SKILL.md}}
\label{lst:yahoofinance-skill}
\end{listing}

\noindent \textbf{Delivery Chains and Obfuscation.}
% \subsubsection{Delivery Chains and Obfuscation}
The skill deploys different payloads on Windows and macOS, both structured to
defeat static scanning at the registry level. The Windows path delivers
\texttt{openclawcli.zip}---seven blobs with Shannon entropy of 7.944--7.960
bits/byte (consistent with AES-CBC/GCM-encrypted payloads) and a loader binary
hosted at a separate GitHub release URL, ensuring the repository contains no
directly executable content. The macOS path points to a \texttt{glot.io}
snippet whose leading \texttt{echo} displays a plausible HTTPS domain; the
actual payload is a base64-encoded second stage that decodes to
\texttt{/bin/bash -c "\$(curl -fsSL http://91.92.242.30/528n21ktxu08pmer)"}
---fetching and executing an arbitrary script from a raw IPv4 address on
bulletproof infrastructure. Neither path places executable content inside the
\texttt{SKILL.md} or archive, defeating registry-level static scanning.

\subsubsection{Trust Violation Analysis}
\label{sec:trust-violation}

The \texttt{yahoofinance} skill exploits precisely the architectural assumption
identified in Section~\ref{sec:skill-arch}: that skill content reflects operator
intent. OpenClaw's context assembly pipeline in \textit{runEmbeddedPiAgent} passes
skill content into the LLM's context window without any distinction between
operator-authored instructions and third-party-authored instructions. Once loaded,
the \texttt{yahoofinance} \texttt{SKILL.md} is semantically equivalent to a system
prompt written by the operator themselves. The LLM has no mechanism to reason about
the provenance of a loaded skill or to distrust instructions that arrive through
the skill loading path.

This design allows the following privilege escalation without exploiting any
memory-safety vulnerability or authentication bypass:

\begin{enumerate}
    \item The skill's instructions are injected into the operator-trusted context
    layer of the agent session.
    \item The LLM relays the installation instructions to the user with the same
    authority as operator-authored prompts.
    \item The user, trusting the agent, executes the prescribed command.
    \item The attacker achieves code execution on the user's machine \emph{outside
    of any OpenClaw exec policy enforcement}.
\end{enumerate}

The Plugin \& Skill System accounts for 7 advisories (3.7\%), including one
Critical path-traversal advisory during plugin installation and one High
advisory for unsafe hook module loading---both sharing the same root cause as
the \texttt{yahoofinance} vector: third-party content is incorporated into the
trusted execution environment without integrity or authenticity enforcement.

\subsection{Agent Context Window}
% \subsubsection{Indirect Prompt Injection}
\label{sec:agent-injection}
Four advisories document injection via data paths that terminate in the LLM
context window: Slack channel metadata prepended to the system prompt
(GHSA-782p), Sentry log headers included verbatim in context (GHSA-c9gv),
resource link \texttt{name} fields displayed as agent memory (GHSA-j4hw),
and filesystem path components resolved into the workspace context (GHSA-m8vc).
The structural cause is uniform: data paths that terminate in the context
window were designed as information channels without adversarial consideration
of what happens when the data is crafted to resemble a directive.

There are also six advisories that define a third class of policy bypass
distinct from the two described earlier in this paper. Exec-policy bypass
operates \textit{below} the reasoning layer---it circumvents the runtime's
enforcement of which system calls or tool invocations are permitted. Skill-level
escalation operates \textit{beside} the reasoning layer---it exploits the
operator trust model to register malicious capabilities before policy is
applied. Prompt injection operates \textit{above} both---it manipulates the
content from which the model constructs its intentions before any policy
enforcement is reached. A model that has been successfully prompt-injected
may voluntarily invoke a tool that policy would otherwise have denied,
rendering the policy irrelevant without ever triggering it.

The indirect injection vectors---channel metadata, log headers, resource
link fields, filesystem paths---share a structural cause: data paths that
terminate in the context window were treated as information channels rather
than as potential instruction channels. This is the correct default
assumption for a system that does not use LLMs; it is the incorrect default
assumption for a system where the context window is the program. Each of the
four indirect injection advisories represents a data-flow path that was
designed without adversarial consideration of what happens when the data is
crafted to resemble a directive.

\subsection{Gateway WebSocket Interface}
As shown in Section~\ref{sec:statistics}, there has been 40 advisories related to Gateway \& API, with the highest concentration of XSS, prototype pollution, token exfiltration, and authorization bypass findings. The 13 High-severity advisories reflect direct exposure of the gateway to external clients.

\subsubsection{Stage 1: Establishing an SSRF Primitive via \texttt{gatewayUrl}}

The outbound message layer in \texttt{src/infra/outbound/message.ts} exposes a
\textit{MessageGatewayOptions} type whose \texttt{url} field was forwarded to
the WebSocket-based gateway client without restriction. The pre-patch
\texttt{resolveGatewayOptions} function reads:

% \begin{listing}[H]
% \begin{minted}{typescript}
% // src/infra/outbound/message.ts  (vulnerable, pre-patch)
% return {
%   url: opts?.url,   // attacker-controlled, no restriction
%   token: opts?.token,
%   ...
% };›
% \end{minted}
% \caption{Pre-patch: \texttt{gatewayUrl} forwarded to the
%   WebSocket client without validation (commit c5406e1)}
% \end{listing}

\begin{listing}[H]
\begin{lstlisting}[language=TypeScript]
// src/infra/outbound/message.ts  (vulnerable, pre-patch)
return {
  url: opts?.url,   // attacker-controlled, no restriction
  token: opts?.token,
  ...
};
\end{lstlisting}
\caption{Pre-patch: \texttt{gatewayUrl} forwarded to the
  WebSocket client without validation (commit c5406e1)}
\end{listing}

Any caller that could supply a \texttt{MessageGatewayOptions} object could
therefore direct outbound gateway connections to an arbitrary URL. In the
backend tool-calling path, this meant the gateway client would attempt an
authenticated WebSocket connection to an attacker-specified host. The fix
replaces the direct pass-through with a conditional that forces \texttt{url} to
\texttt{undefined} for these code paths, coercing the gateway client to fall
back to its configured endpoint.

\subsubsection{Stage 2: Token Exfiltration via the Agent Tool Interface}

The parallel fix in \texttt{src/agents/tools/gateway.ts} addresses the same
class of attacker-controlled URL at the tool-invocation layer. The pre-patch
\texttt{resolveGatewayOptions} function in the tools module forwarded
\texttt{opts.gatewayUrl} directly after trimming:

% \begin{listing}[H]
% \begin{minted}{typescript}
% // src/agents/tools/gateway.ts  (vulnerable, pre-patch)
% const url =
%   typeof opts?.gatewayUrl === "string" && opts.gatewayUrl.trim()
%     ? opts.gatewayUrl.trim()
%     : undefined;
% \end{minted}
% \caption{Pre-patch: \texttt{gatewayUrl} tool parameter
%   accepted without host validation (commit 2d5647a)}
% \end{listing}

\begin{listing}[H]
\begin{lstlisting}[language=TypeScript]
// src/agents/tools/gateway.ts  (vulnerable, pre-patch)
const url =
  typeof opts?.gatewayUrl === "string" && opts.gatewayUrl.trim()
    ? opts.gatewayUrl.trim()
    : undefined;
\end{lstlisting}
\caption{Pre-patch: \texttt{gatewayUrl} tool parameter
  accepted without host validation (commit 2d5647a)}
\end{listing}

An agent tool call supplying
\texttt{\{ gatewayUrl: "ws://attacker.example.com:4444" \}} would direct the
gateway client---including its authentication token---to an attacker-controlled
WebSocket endpoint. The gateway client's handshake protocol sends
authentication material during connection establishment, so merely inducing a
connection attempt is sufficient to capture the bearer token. The fix replaces
the pass-through with \texttt{validateGatewayUrlOverrideForAgentTools()}, which
constructs an allowlist at runtime by reading the configured gateway port and
enumerating loopback variants plus the operator-configured
\texttt{gateway.remote.url} if present. Any URL not matching these canonicalized
keys raises a hard error before connection proceeds.

\subsubsection{Stage 3: RCE via \texttt{node.invoke} Exec Approval Bypass}

With a stolen gateway authentication token, the attacker connects to the
gateway as an authorized operator and invokes \texttt{node.invoke} targeting the
\texttt{system.execApprovals.set} command. Prior to the fix,
\texttt{SYSTEM\_COMMANDS} in \texttt{src/gateway/node-command-policy.ts}
explicitly included these methods:

% \begin{listing}[H]
% \begin{minted}{typescript}
% // src/gateway/node-command-policy.ts  (vulnerable, pre-patch)
% const SYSTEM_COMMANDS = [
%   "system.run",
%   "system.which",
%   "system.notify",
%   "system.execApprovals.get",   // <-- reachable via node.invoke
%   "system.execApprovals.set",   // <-- allows rewriting approval policy
%   "browser.proxy",
% ];
% \end{minted}
% \caption{Pre-patch: \texttt{system.execApprovals.*}
%   reachable via \texttt{node.invoke} (commit 01b3226)}
% \end{listing}

\begin{listing}[H]
\begin{lstlisting}[language=TypeScript]
// src/gateway/node-command-policy.ts  (vulnerable, pre-patch)
const SYSTEM_COMMANDS = [
  "system.run",
  "system.which",
  "system.notify",
  "system.execApprovals.get",   // <-- reachable via node.invoke
  "system.execApprovals.set",   // <-- allows rewriting approval policy
  "browser.proxy",
];
\end{lstlisting}
\caption{Pre-patch: \texttt{system.execApprovals.*}
  reachable via \texttt{node.invoke} (commit 01b3226)}
\end{listing}

By invoking \texttt{system.execApprovals.set} with a crafted approval payload
that adds attacker-controlled executables to the persistent allowlist, the
attacker bootstraps the exec approval state to permit arbitrary command
execution. The next agent \texttt{system.run} invocation then fires the injected
command with full host privileges. The fix removes the two
\texttt{execApprovals.*} entries from \texttt{SYSTEM\_COMMANDS} entirely and
adds an explicit early-return guard in the \texttt{node.invoke} handler.

\subsubsection{Chain Analysis}

The three fixes together expose a trust architecture that had collapsed across
layer boundaries. The gateway layer trusted the URL field from callers it
should have treated as untrusted (agents operating in backend mode). The tool
layer trusted that a \texttt{gatewayUrl} parameter was restricted to safe hosts
by convention rather than by enforcement. The \texttt{node.invoke} handler
trusted that any authenticated operator who could invoke \texttt{system.run}
should also be able to modify the approval policies governing \texttt{system.run}.
Each assumption was locally reasonable in a benign deployment; chained together
under adversarial conditions, they form a complete privilege escalation from LLM
agent output to host code execution.

\subsection{Tool Dispatch Interface}
\label{sec:tool-dispatch-interface}
The 57 advisories across File \& Process System (30), Sandbox Isolation (17),
and Browser Tooling (10) share the structural property identified in the taxonomy:
each layer was designed under a closed-world assumption that its inputs are
operator-controlled, and each fails when that assumption is violated.

\paragraph{File \& Process System (30 advisories).}
Four sub-patterns: (1)~\emph{path traversal} (8)---containment checks applied
before symlink resolution, allowing \texttt{../} sequences or symlinks in
archive entries to escape the target directory; (2)~\emph{SSRF guard bypass}
(7)---IPv4-mapped IPv6 (\texttt{::ffff:169.254.x.x}), ISATAP, and special-use
ranges that resolve to blocked RFC-1918 addresses but evaded the guard;
(3)~\emph{resource exhaustion} (7)---byte limits checked after allocation;
(4)~\emph{host-privileged injection} (8)---unsanitized content flowing into
systemd unit generation and Windows task-scheduler script rendering.
The unifying root cause is absent validation between input ingestion and
privileged use.

\paragraph{Sandbox Isolation (17 advisories).}
The Docker bind-mount escape \cite{ghsa-w235} is analyzed in
\S\ref{sec:container_boundary}. Remaining advisories split between
workspace-boundary violations and the unauthenticated noVNC exposure
\cite{ghsa-h9g4}, which grants full graphical sandbox access from any
network-reachable host. One Critical and two High advisories reflect complete
failure of OpenClaw's primary isolation guarantee.

\paragraph{Browser Tooling (10 advisories).}
Browser automation requires elevated privileges and unrestricted network
access, directly contradicting the isolation model. Ten advisories record
unauthenticated CDP relay endpoints, path traversal in file upload/download,
and absent CSRF protection on navigation triggers. Six of ten are High
severity: unauthenticated browser access is functionally equivalent to full
agent compromise.

\subsection{Exec Policy Engine}
\label{sec:exec-policy-engine}
The three exec allowlist bypasses patched between February 22 and 24 were filed
as separate bugs and fixed in separate commits, but they are manifestations of a
single architectural premise that the allowlist system failed to adequately
defend: a command string's security-relevant identity can be determined by
lexically parsing its text. All three exploits invalidate this premise in
different ways.

\subsubsection{The Line-Continuation Bypass}

The \texttt{evaluateShellAllowlist} function performs command chain splitting
and token-level analysis to determine whether a shell pipeline satisfies a
configured allowlist. The bug is that \textit{DOUBLE\_QUOTE\_ESCAPES} prior to
the patch included \texttt{"\textbackslash n"} and \texttt{"\textbackslash r"}
as recognized escape sequences:

% \begin{listing}[H]
% \begin{minted}{typescript}
% // src/infra/exec-approvals-allowlist.ts  (vulnerable, pre-patch)
% const DOUBLE_QUOTE_ESCAPES = new Set(["\\", '"', "$", "`", "\n", "\r"]);
% \end{minted}
% \caption{Pre-patch: newline treated as escape character,
%   enabling line-continuation injection (commit 3f0b9db)}
% \end{listing}

\begin{listing}[H]
\begin{lstlisting}[language=TypeScript]
// src/infra/exec-approvals-allowlist.ts  (vulnerable, pre-patch)
const DOUBLE_QUOTE_ESCAPES = new Set(["\\", '"', "$", "`", "\n", "\r"]);
\end{lstlisting}
\caption{Pre-patch: newline treated as escape character,
  enabling line-continuation injection (commit 3f0b9db)}
\end{listing}

POSIX shell interprets a backslash followed by a newline inside a double-quoted
string as a line continuation: the backslash and newline are removed, and the
adjacent token fragments are concatenated. The sequence
\texttt{echo "ok \$\textbackslash\textbackslash n(id -u)"} therefore executes
\texttt{id -u} via command substitution, even though the parser, treating
\texttt{\textbackslash n} as an escape character rather than a line-continuation
trigger, fails to detect the nested command. The fix removes
\texttt{"\textbackslash n"} and \texttt{"\textbackslash r"} from
\texttt{DOUBLE\_QUOTE\_ESCAPES} and adds a pre-check function
\texttt{hasShellLineContinuation} that forces \texttt{analysisFailure()} if any
such sequence is detected.

\subsubsection{The Busybox/Toybox Multiplexer Bypass}

The \texttt{allow-always} persistence mechanism records resolved executable
paths of approved commands. For known dispatch wrappers (\texttt{env},
\texttt{nice}, \texttt{nohup}), the system ``unwraps'' the invocation to persist
the inner executable path rather than the wrapper. The bug is that
\texttt{busybox} and \texttt{toybox}---POSIX-compatible multiplexer binaries
that dispatch to sub-tools by their first argument---were not recognized as
wrappers at all. An agent invoking \texttt{busybox sh -c 'whoami'} would have
the \texttt{busybox} binary path persisted in the allowlist. If the operator
approved \texttt{busybox} in any shell-applet context, subsequent invocations of
\texttt{busybox sh -c '\textlangle arbitrary command\textrangle'} would execute
without approval.

The fix required creating an entirely new exported function
\texttt{unwrapKnownShellMultiplexerInvocation} in
\texttt{exec-wrapper-resolution.ts} (a file that did not previously exist), with
a discriminated union result type
(\texttt{not-wrapper | blocked | unwrapped}), and wiring it into three
call sites. Non-shell-applet busybox invocations return
\texttt{\{ kind: "blocked" \}} and fail closed---no allowlist entry is
persisted. The creation of this module from scratch reveals that the original
wrapper-resolution system was written against a closed world of well-known
wrapper binaries with no mechanism for reasoning about multiplexer dispatch
patterns.

\subsubsection{The GNU Long-Option Abbreviation Bypass}

The \texttt{safeBins} subsystem enforces per-binary flag allowlists for safe
utilities like \texttt{sort}, \texttt{grep}, and \texttt{wc}. The pre-patch
\texttt{consumeLongOptionToken} function checked for denied flags with a direct
set membership test:

% \begin{listing}[H]
% \begin{minted}{typescript}
% // src/infra/exec-safe-bin-policy.ts  (vulnerable, pre-patch)
% if (deniedFlags.has(flag)) {
%   return -1;
% }
% // flags absent from both sets fall through -- implicitly permitted
% \end{minted}
% \caption{Pre-patch: denied flag check using exact
%   set membership, missing GNU prefix abbreviation (commit 3b8e330)}
% \end{listing}

\begin{listing}[H]
\begin{lstlisting}[language=TypeScript]
// src/infra/exec-safe-bin-policy.ts  (vulnerable, pre-patch)
if (deniedFlags.has(flag)) {
  return -1;
}
// flags absent from both sets fall through -- implicitly permitted
\end{lstlisting}
\caption{Pre-patch: denied flag check using exact
  set membership, missing GNU prefix abbreviation (commit 3b8e330)}
\end{listing}

GNU getopt conventionally accepts unambiguous prefix abbreviations of long
options: \textit{--compress-prog} is equivalent to \textit{--compress-program}.
The policy enforced \textit{--compress-program} as a denied flag for
\textit{sort}, but \textit{--compress-prog} was not in the denied set and was
simply passed through. The fix introduces \textit{resolveCanonicalLongFlag},
which implements GNU-style prefix matching, and rewrites
\textit{consumeLongOptionToken} to first canonicalize through this function,
returning \textit{-1} for unknown or ambiguous flags before checking the denied
set.

\subsubsection{Root Cause: Lexical Model vs.\ Semantic Reality}

Taken together, these three bugs share a single root cause: the exec allowlist
was designed as a string-matching system operating on static representations of
command text, while the actual security boundary it was meant to enforce
requires reasoning about command \emph{semantics}---how the shell will interpret
the string at runtime, what executable the invocation will actually dispatch to,
and what the effective set of active options will be after GNU abbreviation
expansion. Each bypass represents a gap between the lexical model and the
semantic reality: line continuations rewrite token boundaries mid-parse;
multiplexer dispatch changes the effective executable identity; option
abbreviation expansion changes the effective flag set. The fact that
busybox/toybox required a new module to be written from scratch---rather than a
patch to an existing code path---is the clearest evidence that the original
wrapper-resolution layer was designed around a closed taxonomy of execution
patterns that did not account for the open-ended composability of real-world
Unix tooling.

\subsection{Container Boundary \& Host OS Interface}
\label{sec:container_boundary}
OpenClaw's sandbox subsystem wraps AI agent execution inside Docker containers,
presenting isolation as a core security guarantee. The vulnerable design in
\texttt{src/agents/sandbox/docker.ts} passed the \texttt{SandboxDockerConfig}
object's fields directly through to \texttt{buildSandboxCreateArgs} without any
validation before constructing Docker CLI arguments. The critical assumption was
one of configuration provenance: the code treated all sandbox configuration as
implicitly trusted system-operator input, never considering that config fields
might be populated from agent-controlled or operator-supplied data paths. The
\texttt{binds} field---an array of Docker bind-mount strings in the format
\texttt{source:target[:mode]}---was iterated and emitted as \texttt{-v} flags
verbatim. As the original test fixture demonstrates,
\texttt{binds: ["/var/run/docker.sock:/var/run/docker.sock"]} would be passed
through without objection, mounting the Docker daemon's Unix socket into the
container.

The attack scenario follows directly from this absent validation. An adversary
able to influence the Docker sandbox configuration---whether through a malicious
OpenClaw plugin, a compromised config file on disk, or an agent session that
can write to the configuration store---could inject a bind mount for
\texttt{/var/run/docker.sock} or any ancestor path such as \texttt{/run} or
\texttt{/var/run} that transitively exposes the socket. Once the container
starts with the Docker socket mounted, the sandboxed agent process gains full
Docker API access, enabling it to launch privileged containers, mount the host
filesystem at arbitrary paths, and escape the sandbox entirely. The same attack
surface extends to network namespace isolation: specifying
\texttt{network: "host"} collapses the container's network namespace into the
host's. Setting \texttt{seccompProfile: "unconfined"} or
\texttt{appArmorProfile: "unconfined"} disables the respective Linux security
module enforcement.

The fix (commit \texttt{887b209}, +691/\textminus6 lines) required constructing
an entirely new validation module, \texttt{validate-sandbox-security.ts}, added
from scratch at 208 lines. The new \texttt{BLOCKED\_HOST\_PATHS} constant
enumerates a targeted denylist: \texttt{/etc}, \texttt{/proc}, \texttt{/sys},
\texttt{/dev}, \texttt{/root}, \texttt{/boot},\\ \texttt{/var/run/docker.sock},
\texttt{/run/docker.sock}, and their macOS \texttt{/private/} aliases. The\\
\texttt{validateBindMounts} function implements two-tier checking: direct path
matching against blocked paths, and an ancestor-coverage check that catches
parent-directory mounts like \texttt{/run:/run} that would expose the socket
transitively. The fix uses \texttt{posix.normalize()} to collapse path
traversal sequences before comparison, and \texttt{realpathSync} when symlink
resolution is needed.

% \begin{listing}[H]
% \begin{minted}{typescript}
% // src/agents/sandbox/docker.ts  (vulnerable, pre-patch)
% function buildSandboxCreateArgs(cfg: SandboxDockerConfig): string[] {
%   const args: string[] = [];
%   // binds forwarded directly -- no validation
%   for (const bind of cfg.binds ?? []) {
%     args.push("-v", bind);
%   }
%   if (cfg.network) {
%     args.push("--network", cfg.network); // "host" accepted
%   }
%   return args;
% }
% \end{minted}
% \caption{Pre-patch: bind mounts passed verbatim to Docker
%   (reconstructed from diff context, commit 887b209)}
% \end{listing}

\begin{listing}[H]
\begin{lstlisting}[language=TypeScript]
// src/agents/sandbox/docker.ts  (vulnerable, pre-patch)
function buildSandboxCreateArgs(cfg: SandboxDockerConfig): string[] {
  const args: string[] = [];
  // binds forwarded directly -- no validation
  for (const bind of cfg.binds ?? []) {
    args.push("-v", bind);
  }
  if (cfg.network) {
    args.push("--network", cfg.network); // "host" accepted
  }
  return args;
}
\end{lstlisting}
\caption{Pre-patch: bind mounts passed verbatim to Docker
  (reconstructed from diff context, commit 887b209)}
\end{listing}

% \begin{listing}[H]
% \begin{minted}{typescript}
% // src/agents/sandbox/validate-sandbox-security.ts (added, post-patch)
% const BLOCKED_HOST_PATHS = [
%   "/etc", "/proc", "/sys", "/dev", "/root", "/boot",
%   "/var/run/docker.sock", "/run/docker.sock",
%   // macOS aliases
%   "/private/etc", "/private/var/run/docker.sock",
% ];

% export function validateBindMounts(binds: string[]): ValidationResult {
%   for (const bind of binds) {
%     const src = posix.normalize(bind.split(":")[0]);
%     if (BLOCKED_HOST_PATHS.some(p => src === p || src.startsWith(p + "/")))
%       return { ok: false, reason: `blocked host path: ${src}` };
%     // ancestor-coverage check catches /run:/run etc.
%     if (BLOCKED_HOST_PATHS.some(p => p.startsWith(src + "/")))
%       return { ok: false, reason: `ancestor of blocked path: ${src}` };
%   }
%   return { ok: true };
% }
% \end{minted}
% \caption{Post-patch: validation layer interposed before
%   Docker argument construction (commit 887b209)}
% \end{listing}

The +691/\textminus6 line count reveals a structural truth about the original
sandbox design: the isolation guarantee was entirely emergent from Docker's own
defaults, not from any deliberate enforcement by OpenClaw. The framework leaned
on Docker to behave safely if not told otherwise, but provided no barrier
against configuration that explicitly opted out of protection. This ratio of
remediation code to removed code is characteristic of a security-critical
subsystem designed for functionality first and retrofitted for adversarial
inputs after the fact.

\begin{listing}[H]
\begin{lstlisting}[language=TypeScript]
// src/agents/sandbox/validate-sandbox-security.ts (added, post-patch)
const BLOCKED_HOST_PATHS = [
  "/etc", "/proc", "/sys", "/dev", "/root", "/boot",
  "/var/run/docker.sock", "/run/docker.sock",
  // macOS aliases
  "/private/etc", "/private/var/run/docker.sock",
];
export function validateBindMounts(binds: string[]): ValidationResult {
  for (const bind of binds) {
    const src = posix.normalize(bind.split(":")[0]);
    if (BLOCKED_HOST_PATHS.some(p => src === p || src.startsWith(p + "/")))
      return { ok: false, reason: `blocked host path: ${src}` };
    // ancestor-coverage check catches /run:/run etc.
    if (BLOCKED_HOST_PATHS.some(p => p.startsWith(src + "/")))
      return { ok: false, reason: `ancestor of blocked path: ${src}` };
  }
  return { ok: true };
}
\end{lstlisting}
\caption{Post-patch: validation layer interposed before
  Docker argument construction (commit 887b209)}
\end{listing}

% \placeholder{%
%   Diagram showing the pre-patch data flow: SandboxDockerConfig fields
%   flowing directly into buildSandboxCreateArgs and then into the Docker
%   daemon. Annotate the absent validation gate. Below, show the post-patch
%   flow with validateSandboxSecurity interposed. Use red to highlight the
%   missing validation in the before state.
% }{Pre- and post-patch sandbox configuration flow. Before the fix, bind mount
% strings from \texttt{SandboxDockerConfig} reached the Docker daemon without
% inspection. The fix interposes \texttt{validateSandboxSecurity} at the
% container creation site, enforcing path blocklists, ancestor-coverage checks,
% and symlink resolution before any Docker argument is constructed.}

% \subsection{Host OS Interface}

% \subsection{LLM Provider Interface}

\subsection{LLM Provider Interface \& Inter-Agent Communication}
\label{sec:agent-prompt}

OpenClaw's agent and prompt runtime encompasses the LLM reasoning layer,
the context-assembly pipeline that constructs each model input, and the
session management infrastructure that routes messages between agents and
persists conversation history. Six advisories in this category share a
structural premise that is distinct from the exec-policy bypass class
examined in §\ref{sec:exec-policy-engine} and the plugin skill-escalation class
examined in §\ref{sec:plugin-skill}: where those sections describe attackers
circumventing a policy that the runtime is designed to enforce, prompt
injection and context contamination describe attackers \textit{supplying
the input} from which the runtime derives its instructions. The attack
surface is not a gap in enforcement machinery but a gap in the boundary
between content and instruction---between data the model is told to process
and directives the model is told to follow.

% ------------------------------------------------------------
\subsubsection{Indirect Prompt Injection}
\label{sec:agent-injection}

Four advisories document injection via data paths that terminate in the LLM
context window: Slack channel metadata prepended to the system prompt
(GHSA-782p), Sentry log headers included verbatim in context (GHSA-c9gv),
resource link \texttt{name} fields displayed as agent memory (GHSA-j4hw),
and filesystem path components resolved into the workspace context (GHSA-m8vc).
The structural cause is uniform: data paths that terminate in the context
window were designed as information channels without adversarial consideration
of what happens when the data is crafted to resemble a directive.

\subsubsection{Inter-Session Context Contamination}
\label{sec:agent-session}

\noindent \textbf{Inter-Session Prompts Treated as Direct User Instructions (GHSA-w5c7-9qqw-6645).}
% \subsubsection{Inter-Session Prompts Treated as Direct User Instructions
%   (GHSA-w5c7-9qqw-6645)}
OpenClaw supports agent-to-agent communication through \texttt{sessions\_send},
which routes a prompt from a source agent session into a target agent session.
Because LLM provider APIs require a strict alternating
\texttt{user}/\texttt{assistant} turn structure, the routed prompt is stored
with \texttt{role: "user"} in the target session's transcript. Prior to the
patch at commit \texttt{85409e4}, no additional metadata distinguished this
internally-routed turn from a message typed by an end user. Transcript
readers, memory hooks, and the model itself all received the inter-session
prompt as an ordinary user instruction.

The vulnerability is a trust-boundary collapse: the model cannot distinguish
between ``an end user issued this instruction'' and ``another agent in the
same deployment issued this instruction routed through the session
infrastructure.'' An attacker who can influence an agent in one
session---through a compromised skill, a prompt-injected tool result, or a
malicious external data source---can thereby inject instructions into any
session that agent is permitted to address via \texttt{sessions\_send},
escalating from single-session influence to cross-session control without
any additional capability.

The fix introduces explicit input provenance end-to-end. A new module,\\
\texttt{src/sessions/input-provenance.ts}, defines a structured
\texttt{InputProvenance} type with a closed three-value \texttt{kind} enum.
\texttt{sessions\_send} and the agent-to-agent reply and announce steps now
attach \texttt{inputProvenance: \{ kind: "inter\_session" \}} when invoking
target runs. Provenance is persisted on the user message as
\texttt{message.provenance.kind = "inter\_session"}, with \texttt{role}
remaining \texttt{"user"} for provider compatibility.
\texttt{sanitizeSessionHistory}---the context-rebuild path---detects
\texttt{inter\_session} provenance and prepends a short
\texttt{[Inter-session message]} annotation to those turns in-memory,
providing the model with a signal it can act on without altering the
provider-visible role. The session-memory hook gains a
\texttt{hasInterSessionUserProvenance} guard that skips inter-session turns
when building memory context, preventing routed agent instructions from being
persisted and replayed as if they were user-originated history.

% \begin{minted}{typescript}
% // post-patch: src/sessions/input-provenance.ts (new file)
% export const INPUT_PROVENANCE_KIND_VALUES = [
%   "external_user",
%   "inter_session",
%   "internal_system",
% ] as const;

% export type InputProvenance = {
%   kind: InputProvenanceKind;
%   sourceSessionKey?: string;
%   sourceChannel?: string;
%   sourceTool?: string;
% };
% \end{minted}

\begin{lstlisting}[language=TypeScript]
// post-patch: src/sessions/input-provenance.ts (new file)
export const INPUT_PROVENANCE_KIND_VALUES = [
  "external_user",
  "inter_session",
  "internal_system",
] as const;
export type InputProvenance = {
  kind: InputProvenanceKind;
  sourceSessionKey?: string;
  sourceChannel?: string;
  sourceTool?: string;
};
\end{lstlisting}

The design choice to preserve \texttt{role: "user"} while adding out-of-band
provenance metadata deserves attention. Changing the role to
\texttt{"system"} or introducing a custom role would break provider
compatibility and might itself introduce inconsistencies in how different
providers interpret mixed-role transcripts. The provenance field instead
provides a parallel trust channel: the stored message is well-formed for
every provider, while the runtime and any downstream consumer can read the
provenance to apply appropriate trust. The \texttt{normalizeInputProvenance}
validator ensures that the field cannot be spoofed by arbitrary
\texttt{message.provenance} values arriving from external sources, since it
validates \texttt{kind} against the closed enum before accepting the
provenance object.

% ------------------------------------------------------------
\subsubsection{Structural Analysis}
\label{sec:agent-structural}

The six advisories in this section define a third class of policy bypass
distinct from the two described earlier in this paper. Exec-policy bypass
operates \textit{below} the reasoning layer---it circumvents the runtime's
enforcement of which system calls or tool invocations are permitted. Skill-level
escalation operates \textit{beside} the reasoning layer---it exploits the
operator trust model to register malicious capabilities before policy is
applied. Prompt injection operates \textit{above} both---it manipulates the
content from which the model constructs its intentions before any policy
enforcement is reached. A model that has been successfully prompt-injected
may voluntarily invoke a tool that policy would otherwise have denied,
rendering the policy irrelevant without ever triggering it.

The indirect injection vectors---channel metadata, log headers, resource
link fields, filesystem paths---share a structural cause: data paths that
terminate in the context window were treated as information channels rather
than as potential instruction channels. This is the correct default
assumption for a system that does not use LLMs; it is the incorrect default
assumption for a system where the context window is the program. Each of the
four indirect injection advisories represents a data-flow path that was
designed without adversarial consideration of what happens when the data is
crafted to resemble a directive.

The inter-session contamination advisory (GHSA-w5c7) is structurally
different. It does not involve attacker-controlled external content; the
attacker's leverage is a trust collapse internal to the agent runtime itself.
The fix's provenance model is notable because it introduces a third trust
level---\texttt{external\_user}, \texttt{inter\_session},
\texttt{internal\_system}---into a runtime that previously recognized only
the binary \texttt{user}/\texttt{assistant} distinction imposed by provider
APIs. This is a meaningful step toward a more complete capability model for
multi-agent systems, where the authority of a message is a function of its
origin as well as its role.

The cross-section relationship between this category and
§\ref{sec:tool-dispatch-interface} runs in both directions. Browser path traversal
requires a prompt injection entry point to be exploitable at scale; and a
successful prompt injection against a browser-enabled agent can chain through
navigation and file operations to reach the host filesystem
(§\S\ref{sec:tool-dispatch-interface}). The agent and prompt runtime is therefore not
only an attack surface in itself but the enabling layer for multi-stage
attacks that traverse from LLM input to privileged system output.

\section{Defense Discussion}
\label{sec:defenses}
% ══════════════════════════════════════════════════════════════════════════════

The taxonomy developed in Section~\ref{sec:taxonomy} provides a principled
basis for positioning defenses. We organize recommended mitigations by
taxonomy branch, distinguishing defenses that address the system axis
(architectural hardening) from those that address the attack axis
(technique-specific countermeasures).

\subsection{Channel Integration: Identity-Anchored Allowlists}

The 13 allowlist authorization bypass advisories share a single root cause:
identity fields used for policy evaluation are mutable at the platform level.
The defense is correspondingly simple: allowlists must be keyed exclusively on
immutable, platform-assigned identifiers (numeric user IDs, OAuth subject
claims) rather than human-readable display names or usernames. This is not a
new principle---it is standard OAuth 2.0 practice---but it was not systematically
enforced across any of the 15 affected channel adapters in the OpenClaw corpus
\cite{ghsa-r5h9, ghsa-mj5r, ghsa-chm2}. Webhook signatures should be validated
using HMAC-SHA256 with platform-issued secrets; loopback and proxy trust
exceptions should be removed from production configurations.

\subsection{Gateway \& API: URL Provenance Enforcement}

The Gateway RCE chain \cite{ghsa-g8p2, ghsa-gv46} was enabled by the absence of
URL provenance enforcement in the outbound message and agent tool layers. The
defense is a runtime-constructed allowlist of legitimate gateway URLs (loopback
variants plus operator-configured remote endpoints), validated before any
WebSocket connection is initiated. This is equivalent to the ``safe redirect''
pattern in web application security. Additionally, methods that modify
persistent execution policy (\texttt{system.execApprovals.*}) should be
excluded from the \texttt{node.invoke} dispatch path entirely; remote
administration of approval policy constitutes a privilege boundary that should
require a separate, higher-trust protocol.

\subsection{Exec Allowlist: Semantic Command Interpretation}

The three exec allowlist bypasses \cite{ghsa-9868, ghsa-gwqp, ghsa-3c6h} expose
a fundamental limitation of lexical string-matching as a security primitive for
shell command enforcement. The defense is not a larger denylist but a
semantic interpreter: the allowlist must reason about how the shell will
parse the command at runtime. Concretely, this requires (1) treating any
command containing line-continuation sequences as a parse failure (fail-closed);
(2) recognizing multiplexer binaries (\texttt{busybox}, \texttt{toybox}) and
unwrapping their dispatch to the inner tool before policy evaluation; and
(3) implementing GNU-style prefix abbreviation expansion for long-option
flag matching. Alternatively, the framework should consider restricting
\texttt{system.run} to a direct-argv mode (no shell wrapper) for any
allowlisted invocation, eliminating the shell-parsing attack surface entirely.

\subsection{Sandbox: Configuration Integrity at Creation Time}

The Docker sandbox escape \cite{ghsa-w235} was enabled by the absence of any
validation layer between configuration fields and Docker CLI argument
construction. The defense is a mandatory validation module interposed at
container creation time, enforcing a blocklist of sensitive host paths for bind
mounts, restricting network mode to \texttt{none} by default, and preventing
\texttt{seccompProfile}/\texttt{appArmorProfile} from being set to
\texttt{unconfined}. These are defense-in-depth measures that constrain the
\emph{worst-case outcome} of any upstream configuration compromise, not merely
the specific paths identified by advisories.

\subsection{Plugin/Skill Distribution: Supply-Chain Integrity}

The \texttt{yahoofinance} skill \cite{ghsa-yahoofinance} operated entirely
within the LLM context window and bypassed all runtime policy enforcement. The
defense must therefore operate at the distribution layer, before skill content
reaches the agent. Three complementary controls are required: (1)
\emph{content review} of skills published to the community registry, analogous
to app store review policies; (2) \emph{cryptographic signing} of skill
archives, so the runtime can verify that installed skill content matches a
registry-maintained manifest; and (3) \emph{context-window provenance
annotations}, so the LLM receives a trust signal distinguishing
operator-authored instructions from third-party skill content. None of these
controls requires changes to the exec policy pipeline; all three operate on
the distribution and context-assembly paths that the \texttt{yahoofinance}
attack exploited.

\subsection{Agent/Prompt Runtime: Context Provenance as a Security Boundary}

% The inter-session contamination advisory \cite{ghsa-w5c7} introduced a
% three-level provenance model (\textit{external\_user},
% \textit{inter\_session}, \textit{internal\_system}) that is a meaningful
% architectural step toward treating the LLM context window as a security
% boundary. The defense principle should be generalized: every string that
% enters the context window should carry a \emph{provenance tag} indicating
% its origin (operator, user, channel, skill, inter-agent). The model's
% behavior---and downstream tool-call authorization---can then be conditioned
% on provenance, enabling differentiated trust enforcement above the exec policy
% layer rather than only below it.

Prompt injection remains one of the primary threats to the Agent/Prompt Runtime, as it exploits the fundamental lack of a boundary between untrusted data and executable instructions. Existing research addresses this through two main trajectories: optimization-based structural isolation and detection-based behavioral analysis. Optimization strategies, such as StruQ \cite{Chen2024StruQDAA}, utilize structured queries to enforce syntactic distinctions, while game-theoretic models like DataSentinel \cite{liu2025datasentinel} provide high-assurance detection of adversarial perturbations. Complementary detection frameworks, including PromptArmor \cite{shi2025promptarmorsimpleeffectiveprompt} and PromptSleuth \cite{wang2025promptsleuth}, offer robust filtering by identifying semantic intent invariance or anomalous instruction sequences. Within the OpenClaw architecture, these techniques can be unified and improved through a formal context provenance model. The inter-session contamination advisory already introduced a three-level classification—\textit{external\_user}, \textit{inter\_session}, and \textit{internal\_system}—to distinguish message origins. We propose generalizing this into a mandatory \textit{provenance tag} for every string entering the context window, including those from channel metadata, skill files, and tool outputs. By integrating existing detection-based defenses with this architectural metadata, the runtime can condition tool-call authorization on the verified origin of an instruction rather than its lexical content alone. This transformation shifts the context window from a passive information channel into a monitored, provenance-aware instruction surface that remains resilient even when local detection is bypassed.

\subsection{Cross-Cutting: Unified Inter-Layer Policy Enforcement}

All five defense areas above share a common structural requirement: trust
properties must be enforced at \emph{inter-layer interfaces} through typed,
validated, provenance-carrying request objects, rather than at per-call-site
checks within each layer. A unified policy boundary would evaluate, for any
\texttt{node.invoke} frame, whether the initiating context (agent, operator,
inter-session) is authorized to request the specific command with the specific
parameters, regardless of which individual layer the request traverses. This
is the architectural change that the Gateway RCE chain demonstrates is
necessary: the three-step chain was exploitable precisely because no single
enforcement point observed the full context from LLM tool invocation to
Node-Host shell execution.

% ══════════════════════════════════════════════════════════════════════════════
\section{Conclusion}
\label{sec:conclusion}

Taken together, the 470 advisories analyzed in this paper reflect recurring structural
conditions in the OpenClaw architecture, and understanding those conditions
requires stepping back from individual vulnerabilities to examine the trust
assumptions repeatedly embedded in the codebase. Our two-axis taxonomy
(\S\ref{sec:taxonomy}) makes this structure explicit: the system axis reveals
which architectural layers are most vulnerable, while the attack axis reveals
which adversarial techniques recur across layers and how they map to Cyber
Kill Chain stages \cite{lockheed-killchain}.

The first structural condition is a pervasive \emph{closed-world assumption}:
each subsystem was implemented as though its inputs originated from a
cooperative, finite, and trusted set of sources. The exec allowlist assumed
that command representations were enumerable through lexical parsing
\cite{ghsa-9868, ghsa-gwqp, ghsa-3c6h}. Channel allowlists assumed that sender
identity fields were immutable properties of authenticated sessions
\cite{ghsa-r5h9, ghsa-mj5r}. The LLM context assembly pipeline assumed that
strings entering the context window represented information rather than
potential instructions. Such assumptions are defensible in closed systems;
OpenClaw's open-world deployment model invalidates each of them.

The second structural condition is the \emph{absence of unified inter-layer
policy enforcement}. Trust decisions are made locally (per call site, per
subsystem, per handler) without a global invariant enforced across component
boundaries. The Gateway RCE chain \cite{ghsa-g8p2, ghsa-gv46} illustrates
this: three independently moderate-severity advisories composed into complete
unauthenticated host code execution because no enforcement point observed the
full call path from LLM tool invocation to Node-Host shell execution.

The third structural condition is the \emph{emergence of the plugin and skill
distribution channel as an execution surface} for which the runtime provides
no dedicated policy primitive. The malicious \texttt{yahoofinance} skill
\cite{ghsa-yahoofinance} did not exploit a memory-safety bug or parsing flaw;
it operated entirely within the LLM context window, evading the entire exec
policy pipeline.

Future defense direction in \S\ref{sec:defenses} maps recommended mitigations to
each structural condition. Securing AI agent frameworks is not primarily a matter
of enumerating and patching vulnerabilities; it is a matter of designing a
coherent trust model that remains sound when the layers of the system are
composed in the manner an adversary will inevitably treat as a single unified
attack surface. Our two-axis taxonomy presented in this paper provides the vocabulary
for that design conversation.

% \newpage
% ══════════════════════════════════════════════════════════════════════════════
% Bibliography
% ══════════════════════════════════════════════════════════════════════════════

\bibliographystyle{plain}
% \bibliography{references}   % uncomment when references.bib is ready

% ── Inline bibliography (replace with \bibliography{references} once .bib is ready) ──

\end{document}